\documentclass[12pt]{article}

\input epsf
\usepackage{amssymb,amsmath,mathtools,wrapfig,subfigure}
\usepackage[matrix,arrow,curve]{xy}
\usepackage{cite}
\usepackage{graphicx,color}
\usepackage[debug,pageanchor=false]{hyperref}
\definecolor{link}{rgb}{.8,.15,.1}
\hypersetup{colorlinks=true,linkcolor=link,citecolor=link,urlcolor=link,linktocpage}

\makeatletter
\@addtoreset{equation}{section}
\makeatother

\def\rr {{\Bbb R}}

\def\del {\partial}

\def\del {\partial}

\def\vol {\mathrm{vol}}

\begin{document}

\begin{titlepage}

\begin{center}

\vskip .3in \noindent



{\Large \textbf{Pure spinor equations to lift gauged supergravity}}
\bigskip

Dario Rosa and Alessandro Tomasiello\\

\bigskip
  Dipartimento di Fisica, Universit\`a di Milano--Bicocca, I-20126 Milano, Italy\\
and\\
INFN, sezione di Milano--Bicocca,
I-20126 Milano, Italy

\vskip .5in
{\bf Abstract }
\vskip .1in

\end{center}

We rewrite the equations for ten-dimensional supersymmetry in a way formally identical to a necessary and sufficient $G$-structure system in ${\cal N}=2$ gauged supergravity, where all four-dimensional quantities are replaced by combinations of pure spinors and fluxes in the internal space. This provides a way to look for lifts of BPS solutions without having to reduce or even rewrite the ten-dimensional action. In particular this avoids the problem of consistent truncation, and the introduction of unphysical gravitino multiplets.


\noindent

\vfill
\eject

\end{titlepage}

\hypersetup{pageanchor=true}

\tableofcontents

\section{Introduction} 
\label{sec:intro}

In the context of string theory compactifications, four-dimensional effective actions are often a useful way to capture and organize the relevant physics. On the other hand, the reduction procedure is often laborious. Moreover,   sometimes the reduced action misses important subtleties: for example, the truncation of modes that defines it is often ``non-consistent'', in that it misses some equations of motion of the ten-dimensional action. This means that lifting a solution to ten dimensions is not guaranteed to work. 
For ``vacuum'' solutions (namely those of the type ${\rm Mink}_4 \times M_6$ or ${\rm AdS}_4\times M_6$) experience shows that it is sometimes faster to look for solutions directly in ten dimensions. 

In this paper, we sidestep the problems associated with reductions. We demonstrate an alternative approach to lifting four-dimensional BPS solutions to ten dimensions. We consider ten-dimensional type II theories on fibrations 
\begin{equation}\label{eq:ds46}
	ds^{2}_{10} = ds^{2}_4 (x) + ds^{2}_6 (x,y)\ :
\end{equation}
the metric on the internal six-dimensional space $M_6$ (with coordinates $y^m$) is allowed to depend on the coordinates $x^\mu$ of the spacetime $M_4$ (corresponding to varying scalars in four dimensions);\footnote{\label{foot:fibr}The fibration is assumed to be topologically trivial; for this reason, we need not introduce connection terms in (\ref{eq:ds46}), and we can work in the gauge $g_{\mu m}=0$. In many applications $M_4$ is homeomorphic to $\rr^4$, and the fibration is automatically topologically trivial. On the other hand, in other cases (such as in presence of black holes, as we will see shortly) spacetime does have nontrivial topological features, and the fibration may be non-trivial. We will sketch in section \ref{sub:fibr} how our results would be changed in presence of such a non-trivial fibration.

(\ref{eq:ds46}) also sets to zero the so-called warping function $A$, an overall function of the internal coordinates $y^m$ which in this context is not particularly natural; this complication could be easily added to our formalism. For similar reasons, the dilaton $\phi$ will be taken to depend on the spacetime coordinates, but not on the internal directions.} a natural Ansatz is made for the supersymmetry parameters. 
We organize the \emph{ten-dimensional} supersymmetry equations in such a way as to resemble the supersymmetry equations one gets in a four-dimensional ${\cal N}=2$ supergravity. The system includes some equations corresponding to multiplets one usually throws away in reductions to gauged supergravity. 

Let us be more precise by recalling how a reduction to gauged supergravity usually proceeds. As is well known, reducing type II on a Calabi--Yau yields ungauged ${\cal N}=2$ supergravity; internal fluxes then correspond to gauging the theory (see for example \cite{polchinski-strominger,Dall'Agata:2001zh,louis-micu}). The $G$-structure approach suggests that this might be true more generally for ${\rm SU}(3) \times {\rm SU}(3)$ structure manifolds \cite{gurrieri-louis-micu-waldram}. This has been argued for by proceeding in two steps \cite{grana-louis-waldram,grana-louis-waldram2}: first, the ten-dimensional theory is formally rewritten as a four-dimensional action; second, one truncates to a finite set of internal forms. Already in the first step, one needs to set to zero certain modes that would in principle lead to additional gravitino fields, beyond the two one expects for an ${\cal N}=2$ theory; these are best avoided because they would lead to null states without a gauge invariance to gauge them away. (Even a massive gravitino multiplet would probably receive a mass through a super-Higgs effect; this would lead again to the same problem.) This in turn leads to setting to zero also some internal RR fluxes, associated to the ``edge of the Hodge diamond'' (in the Calabi--Yau case they would correspond to cohomologies like $h^{1,0}$, $h^{2,0}$). In the second step, finding an appropriate set of internal forms is in general challenging \cite{kashanipoor-minasian}, although it can be done on coset manifolds \cite{kashanipoor,cassani-kashanipoor}. 

In our approach, we avoid completely the truncation problem, since we are just rewriting the supersymmetry equations in four-dimensional language. We also avoid the gravitino problem: we are not attempting to write an action, but simply rewriting the supersymmetry equations. And indeed we get some equations that appear to be formally associated to the extra gravitinos, and some associated to the ``edge of the diamond'' vector multiplets which are usually set to zero (sections \ref{sub:edge-vm} and \ref{sub:edge-gino} below). 

Of course we are not saying that writing an effective action is not important: it allows for example to obtain the particle content around a particular vacuum. For classical solutions, however, it can be more practical to work directly in ten dimensions. As we mentioned, this is illustrated by vacuum solutions: recent examples of solutions which were found directly in ten dimensions, and for which an effective four-dimensional theory is not known yet, are the vacua \cite{petrini-zaffaroni,ajtz} in massive IIA. 

Let us also stress again that most reductions are not consistent. In other words, in general a solution found in an effective four-dimensional supergravity is not guaranteed to have a lift to ten dimensions. So far this issue has not been too important, because for many notable solutions there are alternative arguments that guarantee this lift. For example, for an ungauged ${\cal N}=2$ theory arising from compactification on a Calabi--Yau, asymptotically Minkowski black holes arise from D-branes or D-brane bound states. For asymptotically AdS black holes the existence of a lift is less obvious (although for example the 1/4 BPS regular black holes in \cite{cacciatori-klemm} can be lifted to M-theory \cite[Sec.~8]{hristov-vandoren}).\footnote{\label{foot:bh}For these black holes the $M_6$ fibration is nontrivial over the horizon. In most of this paper we have assumed that the fibration is topologically trivial (see footnote \ref{foot:fibr}), but the extension to nontrivial fibrations will be discussed in section \ref{sub:fibr}.} Even setting aside the consistent truncation issue, in this context of AdS black holes our results might be helpful to clarify the attractor mechanism \cite{ferrara-kallosh-strominger}. Already for Minkowski black holes, in ten dimensions the attractor equations get reformulated \cite{hmt} in ten dimensions as a flow on the holomorphic three-form $\Omega$ of the internal Calabi--Yau (in the IIB case), and the entropy becomes related to the Hitchin functional \cite[Sec.~5.1]{hitchin-gcy}. Some of our equations look already very similar to attractor equations, and we plan to come back on this topic in the future. 

Here is a sketch of how we derive our lifting equations. First, since our ten-dimensional analysis will be in the language of differential forms, we need a version of the supersymmetry equations in four dimensions written in the language of $G$-structures. There are several analyses of this type, ranging from a classification of solutions in minimal gauged supergravity \cite{caldarelli-klemm} to a recent characterization of solutions in a general gauged theory \cite{meessen-ortin} for the \emph{timelike case}, namely when the Killing isometry $k$ associated to the preserved supercharge is timelike. We extract from \cite{meessen-ortin} a minimal set of equations,\footnote{\label{foot:MO}\cite{meessen-ortin} is concerned with characterizing the metric in terms of a convenient choice of coordinates, as is often done in supergravity in low dimensions, and find it convenient to work with a slightly redundant set of equations; for our purposes, it is important to choose a minimal, non-redundant system.} consisting of the boxed equations (\ref{eq:timesusy}), (\ref{eq:timesusyhyper}), (\ref{eq:timesusygauginooneform}). 

We then need a system in ten dimensions. For vacuum solutions, we know already that preserved supersymmetry is equivalent to the ``pure spinor equations'' \cite{gmpt2} that constrain solutions of the form ${\rm Mink}_4 \times M_6$ or ${\rm AdS}_4 \times M_6$. The pure spinors are differential forms $\phi_\pm$ on $M_6$ that define an ${\rm SU}(3) \times {\rm SU}(3)$ structure on the ``generalized tangent bundle'' $T \oplus T^*$; the supersymmetry equations constrain them so that, for example, for the Minkowski case $M_6$ needs to be a generalized complex manifold \cite{hitchin-gcy}. The use of $T \oplus T^*$ is crucial for these results: it allows to unify different $G$-structures, and leads to more elegant equations than it would otherwise be possible. Using again the geometry of $T \oplus T^*$, then, in \cite{10d} an extension of the pure spinor equations was found to solutions of type II supergravity which are not necessarily factorized as in a vacuum solution.

We specialized the system in \cite{10d} to metrics which are factorized as in (\ref{eq:ds46}) (where, as we said, the spacetime $M_4$ is neither ${\rm Mink}_4$ nor ${\rm AdS}_4$).\footnote{One might think it should also be possible to specialize directly the supersymmetry equations in the original fermionic form, without using \cite{10d}. This is possible in principle, but it turns out to be less convenient. For example, the internal gravitino would give rise to a generalization of the pure spinor equations of \cite{gmpt2}, but one which is much less compact than (\ref{eq:ext}) below.}  After some massage, we were able to cast our ten-dimensional system in very similar terms to four-dimensional supersymmetry. The pure spinor equations for vacua are now promoted to first-order equations that dictate the evolution of the internal geometry with the point in spacetime, and we show how they correspond to the equations in four dimensions dictating the evolution of the scalars of the theory. (In black hole applications, these equations would become the attractor equations mentioned earlier.) Most of our equations can be matched with four-dimensional ones in a way consistent with \cite{grana-louis-waldram,cassani-bilal}. We also have, however, some new conditions, which we identify as originating with the new vector multiplets and with the extra gravitinos we mentioned earlier. 
The system is summarized in (\ref{eq:summary10dtime}).

As an aside, we also look beyond the timelike case in four dimensions, inspired by the manipulations we performed to compare ten and four dimensions. The Killing vector $k$ is, pointwise in $M_4$, generically timelike, but in many interesting cases it is lightlike in some locus, or even everywhere. This latter case is for example what happens for ${\cal N}=1$ vacua, or for domain-wall solutions. Again the ``generalized tangent bundle'' point of view allows to unify the timelike and lightlike case: both correspond to the same structure group on $T \oplus T^*$ (see appendix \ref{app:pure}). However, this time the equations we obtained are rather more complicated than in the timelike case, which is why we have given less prominence to this aspect in this paper.

Here is a detailed plan of this paper. We begin in section \ref{sec:4d} by reviewing some facts about four-dimensional spinors and the $G$-structures they define. While in the lightlike case the two spinors coincide and define an $\rr^2$ structure, in the timelike case they define an identity structure --- or in other words a vielbein.  We will use this in section \ref{sec:4dsugratime}, where we will establish a system in terms of exterior algebra (a subset of the equations presented in \cite{meessen-ortin}) equivalent to preserved supersymmetry in a gauged ${\cal N}=2$ supergravity with an arbitrary number of vector multiplets and hypermultiplets. In section \ref{sec:10d} we will turn to type II theories, and specialize the system in \cite{10d} to a factorized geometry of the form (\ref{eq:ds46}), allowing any possible metric on $M_4$. In section \ref{sec:time10d} we will then add the timelike hypothesis, and use again the natural vielbein introduced in section \ref{sec:4d} to rewrite the equations found in section \ref{sec:10d}. As we already mentioned, they will be naturally reorganized in a way reminiscent of the system found in section \ref{sec:4dsugratime}. In section \ref{sec:4dgen} we return to four dimensions; inspired by our specialization in section \ref{sec:10d} of the ten-dimensional system in \cite{10d}, we will find a system of equations which is equivalent to preserved supersymmetry even without the timelike assumption. Each of these sections has a concluding subsection summarizing its main results. 
Section \ref{sec:conc} contains a summary of the main results of our paper.


\section{Geometry of four-dimensional spinors} 
\label{sec:4d}
 
We begin by reviewing some facts about the geometry defined by four-dimensional spinors. In particular, in section \ref{sub:timelike4d} we show which exterior differentials ((\ref{eq:listtime}) below) are equivalent to the covariant derivatives of two spinors in the timelike case (to be defined in section \ref{sub:4d2eps}). This result will be useful both for section \ref{sec:4dsugratime}, where we consider four-dimensional ${\cal N}=2$ supergravity, and in section \ref{sec:time10d} where we consider type II supergravity. The general case, beyond the timelike assumption, will be considered in section \ref{sec:4dgen}.

\subsection{One spinor} 
\label{sub:4deps}

Let us consider a single four-dimensional spinor $\zeta_+$, of positive chirality. Most of this material was reviewed in \cite{10d} and \cite{ckmtz}.

It will be convenient to work with real gamma matrices. In this basis, the Majorana conjugate of $\zeta_+$ is simply the naive conjugate $(\zeta_+)^*\equiv \zeta_-$; hermitian conjugation acts by
\begin{equation}\label{eq:gdagger}
	\gamma_\mu^\dagger = \gamma^0 \gamma_\mu \gamma^0\ .
\end{equation}
If we also introduce barred spinors $\overline{\zeta_\pm}\equiv \zeta_\pm^\dagger \gamma^0$, we can also define form bilinears; in particular, a one-form (or vector) $k$ and a two-form $\omega$. This can be summarized by saying\footnote{In this paper, $*$ will be the four-dimensional Hodge star operator unless otherwise noted.} 
\begin{equation}\label{eq:bisp}
	\zeta_+ \otimes\overline {\zeta}_+ = k + i * k
	\ ,\qquad
	\zeta_+ \otimes \overline{\zeta}_- \equiv \omega 
	\ .
\end{equation}
The structure of the first bilinear is dictated by the fact that the left action of the chiral $\gamma$ on any form $C$ reads, in four dimensions,\footnote{The generalization to any dimension is described in \cite[App.~A.2]{10d}.}
\begin{equation}\label{eq:gl}
	\gamma C = i * \lambda C \ , \qquad  \lambda C \equiv (-)^{\lfloor \frac12 {\rm deg}(C) \rfloor} C\ .
\end{equation}
It also follows that $* \omega = i \omega$. 

An easy Fierz computation also gives us\footnote{\label{foot:clifford}Throughout the paper, we will confuse forms and bispinors related by the Clifford map $dx^{M_1}\wedge \ldots \wedge dx^{M_k} \leftrightarrow \gamma^{M_1\ldots M_k}$; this is often denoted with a slash, but we will drop it to make our equations more readable.}
\begin{equation}\label{eq:zeps}
	k \, \zeta \equiv  k_\mu \gamma^\mu \zeta_+ = 
	\frac14\gamma^\mu \zeta_+ \overline{\zeta_+} \gamma_\mu \zeta_+ = -\frac12 (1- \gamma) k \zeta_+ = - k \,\zeta_+  
	\qquad \Rightarrow \qquad k \,\zeta_+ = 0 \ ,
\end{equation}
where we have used the well-known formula $\gamma^\mu C_k \gamma_\mu = (-)^k(d-2k) C_k$.

(\ref{eq:zeps}) implies that $k^2=0$. We thus have a map $\psi$ from the vector space $\rr^4$ of Weyl spinors of positive chirality to the cone of null vectors. We can also restrict $\psi$ to the subspace of spinors of fixed norm: 
\begin{equation}\label{eq:hopf4}
	\{ \zeta_+ | \zeta_+^\dagger \zeta_+ = 1\} \cong S^3
	 \ \longrightarrow \ \{ k| k^2=0 \,,\ k_0=1 \}\cong S^2
\end{equation}
This is now nothing but the Hopf fibration map, whose fibre is $S^1 \cong {\rm U}(1)$. It follows that, given $k$, one can determine $\zeta$ up to an overall phase.

(\ref{eq:zeps}) also implies that $[k, \zeta_+ \otimes \overline{\zeta_-}]=0$,  which translated into forms reads $k\wedge \omega=0$. This means in turn that there exists a $w$ such that
\begin{equation}
	\omega = k \wedge w \ ,
\end{equation}
where $w$ is a \emph{complex} one-form, which also annihilates $\zeta_+$. Since we now have
\begin{equation}\label{eq:4dpure}
	k \zeta_+ = w \zeta_+ = 0 \ ,
\end{equation}
$\zeta_+$ is annihilated by two combinations of gamma matrices; in other words, it is a pure spinor. 

One can now also show that
\begin{equation}
	k\cdot w = w^2= 0 = k^2 \ ,\qquad w\cdot\bar w =2\ .
\end{equation}
We can think of $k$ and $w$ as elements of a local frame: $k=e^+$, $w=e_2-i e_3$. We have now exhausted the list of one-forms we can define from $\zeta_+$ alone; we see that a single spinor is not enough to define a vielbein (similarly to the discussion for 10d spinors in \cite[Sec.~2]{10d}). In group theory terms, this is because  $\zeta_+$ has a stabilizer isomorphic to the group of two-dimensional translations $\rr^2$,\footnote{The stabilizer of the light-like vector $k$ is 
${\rm SO}(2)\ltimes \rr^2$; $w$ breaks the SO(2) to the identity. For more details see \cite{10d}. Alternatively, one can compute the stabilizer of $\zeta$ directly. In a vielbein where $k= e^+$, the stabilizer is spanned by $\gamma^{+i}$, where $i\neq -$. These generate the abelian Lie algebra $\rr^2$.} and thus defines an $\rr^2$ structure, rather than an identity structure which would be necessary to define a vielbein.
It is then often convenient to complete the vielbein by introducing an additional null real one-form $e^-$ such that
\begin{equation}\label{eq:e-}
	(e^-)^2=0 \ ,\qquad e^-\cdot k = 2 \ ,\qquad e^- \cdot w =0 \ ,
\end{equation}
as was done in \cite{ckmtz} (and in \cite{10d} in ten dimensions).


\subsection{Two spinors: the timelike and null cases} 
\label{sub:4d2eps}

Since we will deal with ${\cal N}=2$ supergravity, we will also need to study the structure defined by two spinors $\zeta_{1+}$, $\zeta_{2+}$. To make our equations more readable, we will drop the subscript ${}_+$: it will be understood from now on that $\zeta_i$, $i=1,2$ are Weyl spinors of positive chirality. Their Majorana conjugates will then be Weyl spinors of negative chirality and, as usual in $\mathcal{N} = 2$ supergravity, they will be denoted with an upper index $\zeta^i$:
\begin{equation}
	\gamma\zeta_i = \zeta_i \ ,\qquad \gamma\zeta^i = - \zeta^i \ .
\end{equation}
The barred version of the $\zeta_i$ will be denoted by $\overline{\zeta^i}$, since they have opposite chirality; and likewise for their complex conjugates $\overline{\zeta_i}$:
\begin{equation}
	\overline{\zeta^i} \gamma= - \overline{\zeta^i} \ ,\qquad
	\overline{\zeta_i} \gamma= \overline{\zeta_i}\ .
\end{equation}

Each of the two spinors $\zeta_i$ will now define its own one-forms $k_i$, $w_i$, and two-forms $\omega_i = k_i \wedge w_i$, following section \ref{sub:4deps}. However, we are now also able to define mixed bilinears: 
\begin{equation}\label{eq:bisp12}
	\zeta_1 \otimes \overline{\zeta^2} \equiv v + i * v \ ,\qquad
	\zeta_1 \otimes \overline{\zeta_2} \equiv \mu (1+ i \vol) + \omega\ .
\end{equation}
This new $\omega$ satisfies $* \omega = i \omega$, just like the $\omega_i$ associated to the individual $\zeta_i$.

The new vector $v$ is almost entirely fixed by the $k_i$ associated to the individual $\zeta_i$ as in (\ref{eq:bisp}). Indeed, one can show, in a similar way as (\ref{eq:zeps}), 
\begin{equation}\label{eq:viz}
	v \zeta_1 = 0 \ ,\qquad 
	v \zeta_2 = - k_2 \zeta_1 \ ,\qquad
	\bar v \zeta_2 = 0 \ .
\end{equation}
This in turn implies 
\begin{equation}\label{eq:zv}
	v \cdot k_i = 0 \ ,\qquad v^2= 0 \ ,\qquad v \cdot \bar v = -k_1 \cdot k_2 \ .  
\end{equation}
When the $k_i$ are not proportional, these relations determine $v$ up to a phase (as we will see later more explicitly). When the $k_i$ are proportional, however, the ambiguity is greater. In other words, there is no formula for $v$ in terms of the $k_i$ alone. We could only find `asymmetrical' formulas such as
\begin{equation}\label{eq:vzw}
	v = \frac1{2 \mu} (k_2 \cdot k_1 w_1 - k_2 \cdot w_1 k_1 )=
	-\frac1{2\bar \mu} (k_1 \cdot k_2 w_2 - k_1 \cdot w_2 k_2 )  \ .
\end{equation}

We saw in section \ref{sub:4deps} that a single spinor $\zeta$ defines an $\rr^2$ structure. We now see that the structure defined by two spinors $\zeta_i$ depends again on whether they are parallel or not. When they are not parallel, the one-forms
\begin{equation}
	k_1 \ ,\qquad k_2 \ ,\qquad {\rm Re} v \ ,\qquad {\rm Im} v 
\end{equation}
together constitute a vielbein (up to an overall rescaling), thanks to $k_i^2=0$ and to (\ref{eq:zv}). Thus the $\zeta_i$ define an identity structure. On the other hand, when the $\zeta_i$ are parallel their common stabilizer is just the stabilizer of one of them, namely $\rr^2$. 

As we learned around (\ref{eq:hopf4}), one can reconstruct $\zeta$ from its associated null vector $k$, up to a phase. Thus, we can tell whether the $\zeta_i$ are parallel by just looking at whether the $k_i$ are parallel. In what follows, the vector
\begin{equation}\label{eq:K4d}
	k \equiv \frac12 (k_1 + k_2) 
\end{equation}
will play a special role. When the two $\zeta_i$ are not parallel (and thus define an identity structure, and a vielbein), $k$ is the sum of two different lightlike vectors, and thus must be timelike; from now on, we will call this \emph{the timelike case}. On the other hand, when the $\zeta_i$ are proportional, the $k_i$ are also proportional, and $k$ is null; we will then call this \emph{the null case} from now on. 

Another useful indicator of which case we are dealing with is the complex quantity $\mu$ in (\ref{eq:bisp12}). By comparing with (\ref{eq:bisp}), we see that $\mu$ should vanish in the null case (when the $\zeta_i$ are parallel). In fact one can be more precise: 
\begin{equation}\label{eq:mu2}
	-16 |\mu|^2 = (\overline{\zeta^1} \zeta^2)(\overline{\zeta_2}\zeta_1) = 
	{\rm Tr}( \zeta_1 \overline{\zeta^1} \zeta^2 \overline{\zeta_2}) = {\rm Tr}((1+\gamma) k_1 (1-\gamma) k_2 )=
	{\rm Tr}(2 (1+ \gamma) k_1 k_2 )= 8 k_1 \cdot k_2\ .
\end{equation}

As we have seen, in the timelike case there is a natural vielbein, while in the null case there is none. Again to retain full generality, we will find it useful to introduce two null vectors 
\begin{equation}
	e^-_1 \ ,\qquad e^-_2\ ,
\end{equation}
satisfying
\begin{equation}\label{eq:e-i}
	(e^-_i)^2=0 \ ,\qquad e^-_i\cdot k_i = 2 \ ,\qquad e^-_i \cdot w_i =0 \ .
\end{equation}

In the timelike case, $k_1$ and $k_2$ do not coincide and are both null; so we can just take $e^-_1$ proportional to $k_2$, and $e^-_2$ proportional to $k_1$. The proportionality constants can be fixed using (\ref{eq:e-i}):
\begin{equation}\label{eq:e-time}
	e^1_- = - \frac{k_2}{|\mu|^2}\ ,\qquad
	e^2_- = - \frac{k_1}{|\mu|^2}\  \qquad ({\rm timelike\ case}).
\end{equation}
From (\ref{eq:vzw}) and (\ref{eq:mu2}) we then also get
\begin{equation}\label{eq:vtime}
	v= - \bar \mu w_1 = \mu \bar w_2 \qquad ({\rm timelike\ case}).
\end{equation}
In the null case, on the other hand, $k_1$ and $k_2$ are proportional: we have
\begin{equation}\label{eq:spinornull}
  \zeta_2 = \alpha (x)\,\zeta_1
\end{equation}
which leads us to 
\begin{equation}
  k_2 = |\alpha(x)|^2\,k_1\ .
\end{equation} 
Therefore we conclude that there is no natural candidate for the $e^-_i$. On the other hand, there is no need to pick two of them, and we can just as well say 
\begin{equation}
  e_2^- = \frac{e_1^-}{|\alpha|^2}\ \qquad ({\rm null\ case})
\end{equation}
where the proportionality between $e_1^-$ and $e_2^-$ is fixed by requiring
\begin{equation}
  e_i^- \cdot k_i = 2 \,\qquad i= 1, 2\ .
\end{equation}

\subsection{${\rm SU}(2)$-covariant formalism} 
\label{sub:su2}

In the timelike case, it will often be useful to collect the bilinears we introduced in section \ref{sub:4d2eps} in an ${\rm SU}(2)$-covariant fashion. 

We define\footnote{\label{foot:sigma}In our conventions, $\sigma_x^i{}_j$ are the conventional Pauli matrices, while $\sigma_{x\,j}{}^i$ are their transposes; notice that the position of the index $x$ does not play any role. Moreover, $\epsilon_{ij}= \epsilon^{ij}=\left(\begin{smallmatrix}
	0 & 1 \\ -1 & 0
\end{smallmatrix}\right)$. We lower (raise) indices acting from the left (right) with $\epsilon$: so for example $\sigma_{x\,ij}= \epsilon_{ik} \sigma_{x\,j}{}^k$, $\sigma_x^{ij}= \sigma_x^i{}_k \epsilon^{kj}$.}
\begin{equation}\label{eq:SU(2)bilinears}
	\begin{split}
		&\zeta_i \otimes \overline{\zeta^j}= (1+i*) v_i{}^j =  (1+i*)(k \delta_{i}{}^{j} + v^x\,\sigma_{x\,i}{}^j) \ , \\
		&\zeta_i \otimes \overline{\zeta_j} = \mu \epsilon_{ij} (1 + i {\rm vol}) + o_{ij} = \mu \epsilon_{ij} (1 + i {\rm vol}) +  o^x \epsilon_{ik} \sigma_x^k{}_j \ ,  
	\end{split}
\end{equation}
which summarize (\ref{eq:bisp}), (\ref{eq:bisp12}); notice that
\begin{align}\label{eq:vx}
  & v^1 = \mathrm{Re}\,v\ , \qquad v^2 = \mathrm{Im}\,v\ , \qquad v^3 = \frac12 (k_1 - k_2)  \ ,
\end{align}
while the vector $k$ is precisely the same vector which we defined in (\ref{eq:K4d}). 

Many of the properties we saw earlier can be now summarized more quickly. For example, one can find 
\begin{equation}
	v_{(i}{}^j \zeta_{k)} = 0 \ \  \Rightarrow \ \ v_{(i}{}^j v_{k)}{}^l = 0 \ ,
\end{equation}
which summarizes (\ref{eq:zeps}), (\ref{eq:viz}) and (\ref{eq:zv}). This tells us that $k \cdot v_x = 0$, $v_x \cdot v_y = \frac13 \delta_{xy} v_z \cdot v_z$, $k^2=v_z \cdot v_z$. To get the overall normalization, one can instead perform a computation similar to (\ref{eq:mu2}), and obtain
\begin{equation}
	v_i{}^k \bar v^l{}_j = 2 |\mu|^2 \epsilon_{ij} \epsilon^{kl}\ .
\end{equation}
This gives us 
\begin{equation}
	k^2 = - |\mu|^2 \ ,\qquad v_x \cdot v_y = \delta_{xy} |\mu|^2 \ ,\qquad
	 k \cdot v_x = 0 \ .
\end{equation}
This means that 
\begin{equation}\label{eq:vielmu}
	\left\{ e^0\equiv \frac1{|\mu|}k ,\ e^x \equiv\frac1{|\mu|}v^x \right\}
\end{equation}
is a vielbein. 


\subsection{Spinor derivatives in the timelike case, and spin connection} 
\label{sub:timelike4d}

In supergravity we also need to discuss spinorial covariant derivatives $\nabla_\mu \zeta_i$. As usual, such covariant derivatives can be conveniently parameterized in terms of the so-called intrinsic torsions of a $G$-structure. As we saw in section \ref{sub:4d2eps}, our two spinors define an identity structure in the timelike case, and an $\rr^2$ structure in the null case. The timelike case is thus significantly simpler, since the intrinsic torsion is in this case nothing but the spin connection itself. We will discuss this case here, and leave the general case (where one might have null loci somewhere) to section \ref{sub:torsgen}.

In the timelike case, one might then think that the information about $\nabla_\mu \zeta_i$ is completely captured by the covariant derivatives of the vielbein (\ref{eq:vielmu}). One might go to a frame where the $\zeta_i$ are constant, and reconstruct then $\nabla_\mu \zeta_i = \frac14 \omega_\mu^{ab} \gamma_{ab} \zeta_i$ from $\nabla_\mu e^a$ or $d e^a$. This, however, forgets the information about the inner product $\mu=\frac14\overline{\zeta_2} \zeta_1$. To see this more clearly, define
\begin{equation}\label{eq:pqt}
	\nabla_\mu \zeta_i = p_{\mu\,i}{}^j \zeta_j\ .
\end{equation}
The $4\times 2\times 2$ complex components of these $p$ are more than the $4\times 6$ real components of the spin connection $\omega^{ab}$. The mismatch is due to the derivatives of $\mu$. Indeed, we can compute in terms of these $p_i{}^j$ the covariant derivatives of the bilinears $\mu$, $k$ and $v^x$, and hence of the vielbein $\{e^0,\,e^x\}$ from (\ref{eq:vielmu}); from the latter we can get the spin connection. Decomposing $p_i{}^j= p^0 \delta_i{}^j + p^x \sigma_{x\,i}{}^j$, we get
\begin{equation}
	p^0 = \frac{d\mu}{2\mu} \ ,\qquad p^x= \frac12 \left( \omega^{0x}+\frac i2 \epsilon^{xyz}\omega_{yz}\right)\ .
\end{equation}
Hence the components of the $p^x$ correspond to the spin connection, while $p^0$ corresponds to the derivatives of $\mu$. Thus the information in $\nabla_\mu \zeta_i$ is contained in the spin connection and in $d\mu$. The spin connection can be extracted from $de^a$, but since we need $d\mu$ anyway, we might as well use directly $dk$, $d v^x$. 

We conclude then that the information in $\nabla_\mu \zeta_i$ is contained in 
\begin{equation}\label{eq:listtime}
	d \mu \ ,\qquad d k_i \ ,\qquad d v \ . 
\end{equation}



\section{${\cal N}=2$ four-dimensional supergravity: timelike solutions} 
\label{sec:4dsugratime}

In this section, we will reformulate the supersymmetry equations for four-dimensional ${\cal N}=2$ supergravity in terms of differential forms, using what we have learned in section \ref{sec:4d}. We will assume that the solution is timelike, in the sense specified in section \ref{sub:4d2eps}: namely, the Killing vector $k=\frac12 (k_1 + k_2)$ is taken to be timelike. 
This case is `generic': even if, in a given solution, there happen to be subsets where $k^2=0$, these have measure zero. There are, however, also (non-generic) null solutions, where $k^2=0$ everywhere, an important example being ${\cal N}=1$ vacua. For this reason, in section \ref{sec:4dgen} we will also take a look at the general case, giving a set of equations which will be inspired by the corresponding set that we will write in ten dimensions.

The timelike case is also notable in that the differential equations one obtains are much nicer than in the general case, in that they can be formulated only in terms of exterior differentials of spinors bilinears and nothing more. The equations that we will write in this section were already derived in \cite{meessen-ortin}, but we will be able to show that only a subset of their system is actually needed for supersymmetry (see however footnote \ref{foot:MO}). 

After some general comments about ${\cal N}=2$ gauged supergravity in section \ref{sub:4dsusytr}, we will reformulate the conditions for supersymmetry (\ref{eq:4dsusyvar}) in sections \ref{sub:4dgrino}, \ref{sub:4dhyno}, \ref{sub:4dgaugino}, and briefly summarize in section \ref{sub:4dsummary} for the reader's convenience.

\subsection{Supersymmetry equations} 
\label{sub:4dsusytr}

We will start by quickly recalling here some features of four-dimensional gauged ${\cal N}=2$ theories. This section is not meant to be a review of the general formalism, for which the reader may consult for example \cite{abcdffm}. We will follow a notation similar to \cite{meessen-ortin}, which recently applied $G$-structures to the general theory. 

A general ${\cal N}=2$ theory consists of:
\begin{itemize}
	\item a graviton multiplet, which contains the metric $g_{\mu\nu}$, two gravitinos $\psi_{i\mu}$, $i=1,2$, and a vector $A^0_\mu$ (with field-strength $T_{\mu\nu}$);
	\item $n_v$ vector multiplets, which contain vectors $A^a_\mu$ (with field-strength $G^a_{\mu \nu}$), gaugini $\lambda^{ia}$ and complex scalars $t^a$ ($a=1,\ldots,n_v$), parametrizing a special K\"{a}hler manifold ${\cal SK}$;
	\item $n_h$ hypermultiplets, which contain $4n_h$ scalars $q^u$ ($u=1,\ldots, 4n_h$) and $2n_h$ hyperini $\kappa_\alpha$ ($\alpha=1,\ldots,2n_h$); in this case the $q^u$ span a quaternionic manifold ${\cal Q}$, whose vielbein is denoted by $U^{i \alpha}_u$.    
\end{itemize}
The $n_v + 1$ vectors are then usually grouped with the notation $A_\mu^\Lambda$ ($\Lambda= 0,\,\dots,\,n_v$). In gauged supergravity, some of these vectors will gauge some symmetries of the scalar manifolds ${\cal SK}$ and ${\cal Q}$, whose generators will be denoted by $k_\Lambda^a$ and $k_\Lambda^u$ respectively. This means that the vectors will appear in covariant derivatives
\begin{align}
  D t^a = d t^a + g\, A^\Lambda k_{\Lambda}^a \ ,\qquad
  D q^u = d q^u + g \, A^\Lambda k_\Lambda^u\ .
\end{align}
Supersymmetry dictates that the Killing vectors $k_\Lambda^a$ should be generated by momentum maps $P_\Lambda$, and that the $k_\Lambda^u$ by hyper-momentum maps $P_\Lambda^u$. 

We will look for solutions where the fermions are set to zero, so that supersymmetry will be unbroken if and only if the variations of the fermions  $\psi_{i\mu}$, $\lambda^{ia}$, $\kappa_\alpha$ are zero. These read
\begin{subequations}\label{eq:4dsusyvar}
\begin{align}
	\label{eq:4dgrino}
	\delta_\zeta \psi_{i\mu} &= D_\mu \zeta_i + \left(T^+_{\mu\nu} \gamma^\nu \epsilon_{ij} -\frac12 \gamma_\mu S_x\sigma^x_{ij}\right) \zeta^j=0\ ,\\
	\label{eq:4dhyno}
	\delta_\zeta \kappa_\alpha &= i U_{\alpha\,iu} D q^u \zeta^i + N_\alpha^i \zeta_i=0 \ , \\
	\label{eq:4dgaugino}
	\delta_\zeta \lambda^{ia} &= i D t^a \zeta^i + \left( 
	(G^{a+} + W^a ) \epsilon^{ij} + \frac i2 W^{a\,x} \sigma_x^{ij}\right) \zeta_j=0 \ .
\end{align}	
\end{subequations}
Here, one- and two-forms act as bispinors (see footnote \ref{foot:clifford}); for Pauli matrix conventions, see footnote \ref{foot:sigma}.
The quantities $S_x$, $W^a$, $W^a_x$ and $N_\alpha^i$ are related to the gauging data (the Killing vectors on ${\cal SK}$ and ${\cal Q}$ and their (hyper)-momentum maps).
The covariant derivatives act on spinors as
\begin{align}
 & D_\mu \zeta_i = \left(\nabla_\mu + \frac{i}{2} \hat{Q}_\mu \right) \zeta_i + \frac i2 \hat{A}^x_{\;\mu} \sigma_i^{x\;j} \zeta_j  \ ,
\end{align}
where the connection $\hat{A}^{x}_{\;\mu}$ is defined from the SU(2) connection $A^x$ on the quaternionic manifold
\begin{equation}\label{connectionA}
  \hat{A}^x_{\;\mu} \equiv \partial_\mu q^u A^x_{\;u} + g A^{\Lambda}_{\; \mu} P_{\Lambda}^{\;x} \ .
\end{equation}

We will now analyze the geometrical content of (\ref{eq:4dsusyvar}). Unlike what happens in ten dimensions, each of these variations can be analyzed separately. 


\subsection{Gravitino equations} 
\label{sub:4dgrino}

We will deal first with the gravitino equation $\delta \psi_{i\mu}=0$ from (\ref{eq:4dgrino}). 

In general, the gravitino equation is the hardest to analyze in supergravity, since it involves derivatives of the spinors. However, as we saw in section \ref{sub:timelike4d}, in the timelike case the information contained in $\nabla_\mu \zeta_i$ is equivalent to the information contained in the exterior derivatives of $\mu$, $k$ and $v_x$. Hence, the gravitino equation (\ref{eq:4dgrino}) is equivalent to the equations for these quantities that can be computed from it; in the ${\rm SU}(2)$-covariant formalism of section \ref{sub:su2}, they read
\begin{equation}
	\label{eq:timesusy}
\fbox{$
\begin{array}{rl}
		D\mu&= S_x v_x - 2 \iota_k T^+\ ,\\
		d k &= -2 {\rm Re} ( S_x \bar o_x  + 2 \bar\lambda T^+ )\ ,\\
		D v_x &= 2\epsilon_{xyz} {\rm Im} ( \bar S_y  o_z ) \ .
\end{array}$}	
\end{equation}
The twisted external differential $D$ acts as
\begin{equation}\label{twisteddiff4d}
 	D \mu = d \mu + i \hat Q \mu \ ,\qquad
 	D v_x = d v_x + \epsilon_{x y z} \, \hat{A}_y \wedge v_z \ .
\end{equation}
Apart from a few redefinitions, these are the same as (3.1), (3.3), (3.5) in \cite{meessen-ortin}. Their (3.2) and (3.4) can be safely dropped: the system (\ref{eq:timesusy}) is equivalent to the four-dimensional gravitino equation (\ref{eq:4dgrino}) in the timelike case, already as it is (see footnote \ref{foot:MO}). It is particularly pleasing that these equations only involve exterior differentials. 


\subsection{Hyperino equations} 
\label{sub:4dhyno}

We now analyze the content of the hyperino equations (\ref{eq:4dhyno}). 

Since they do not involve any derivatives of the $\zeta_i$, they are easier to understand. Their full geometrical content can be obtained by expanding along an appropriate basis of spinors. Since we are in the timelike case, this basis can be taken to be the $\zeta_i$ themselves. To project the (\ref{eq:4dhyno}) on this basis, we can simply multiply them from the left by $\overline{\zeta_k}$. As we mentioned, the $N_\alpha^i$ in (\ref{eq:4dhyno}) can be derived from the gauging data; the precise formula is 
\begin{equation}
  N_\alpha^i = g\, U_{\alpha j u} \bar{\mathcal{L}}^\Lambda k_\Lambda^{\;u} \epsilon^{ji}\ .
\end{equation} 
We get the equation
\begin{equation}
  U_{\alpha i u} \bigl( i [k\, \delta^{\;i}_{k} + v_x\, \sigma^{x\; i}_k] \cdot Dq^u - g\, \bar{\mathcal{L}}^{\Lambda}k_{\Lambda}^{\;u} \, \mu \,\delta_k^{\;i} \bigr) = 0\,.
\end{equation}
Now, since $U_{\alpha iu}$ is a vielbein on the quaternionic manifold ${\cal Q}$, we have $U^u_{\alpha i} U^{\alpha i}_v = \delta^u_v$. Moreover, the tensors
\begin{equation}
	\Omega^x_u{}^v =  i \sigma^x_i{}^k  U^{\alpha i}_u U_{\alpha k}^v 
\end{equation}
are a triplet of complex structures defined on ${\cal Q}$. Using this, one obtains the single equation
\begin{equation}\label{eq:timesusyhyper}
  \fbox{$i\, k \cdot D q^v + \Omega^{xv}{}_u v_x \cdot D q^u - g\,\bar{\mathcal{L}}^{\Lambda}k_{\Lambda}^{v} \, \mu = 0$} \ .
\end{equation}
This equation already appeared in \cite[Eq.~(3.24)]{meessen-ortin}.

\subsection{Gaugino equations} 
\label{sub:4dgaugino}

It remains to consider the gaugino equations (\ref{eq:4dgaugino}). Just like for the hyperino equations, these do not involve any spinor derivatives; hence, again we can extract their full geometrical meaning by multiplying them from the left by $\overline{\zeta_k}$. This gives
\begin{equation}\label{eq:timesusygaugino}
  i\,(v)_k{}^i \cdot D t^a + \epsilon^{ij} o_{kj} \llcorner G^{a\,+} - \mu W^a \delta_k{}^i - \frac i2 \,\mu W^{a\,x} \sigma^{x\,i}{}_k = 0\, . 
\end{equation}
This appeared in \cite[Eq.~(3.7)]{meessen-ortin}. It is a set of four scalar equations; we can also recast them as a single equation for a $1$-form:
\begin{equation}\label{eq:timesusygauginooneform}
 \fbox{  $2 i \bar{\mu} D t^a - 4 \iota_k G^{a +} + 2 W^a k - i W^{a x} v^x  = 0$ }\ .
\end{equation}
This expression will be particularly useful for our comparison with ten-dimensional supersymmetry in section \ref{sec:time10d}.


\subsection{Summary: four-dimensional timelike case} 
\label{sub:4dsummary}

We have found in this section that preserved supersymmetry is equivalent to the system given by the boxed equations (\ref{eq:timesusy}), (\ref{eq:timesusyhyper}), (\ref{eq:timesusygauginooneform}). These come respectively from the variations of the gravitino, of the hyperinos and of the gauginos. The system is formulated in terms of exterior calculus only, and it does not have any redundancy.



\section{Ten dimensions} 
\label{sec:10d}

We will now consider supersymmetry in ten-dimensional type II supergravity. As anticipated in the introduction, we will specialize the system found in \cite{10d} to a topologically trivial fibration: 
\begin{equation}\label{eq:metAn}
	ds^{2}_{10} = ds^{2}_4 (x) + ds^{2}_6 (x,y)\ .
\end{equation}
$M_4$ is a four-dimensional spacetime with coordinates $x^\mu$ and Lorentzian metric $g_4$, and $M_6$ is a compact space with coordinates $y^m$ and Riemannian metric $g_6$, admitting an ${\rm SU}(3) \times {\rm SU}(3)$ structure. We will not introduce any a priori constraint on either $g_4$ or $g_6$. The extension of our results to fibrations which are topologically non-trivial will be sketched in section \ref{sub:fibr}. 

In section \ref{sec:time10d} we will rewrite this system and compare it to the four-dimensional system presented in section \ref{sec:4dsugratime}. Our results in this section will also help us in section \ref{sec:4dgen}, where we will present a system equivalent to supersymmetry in four-dimensional ${\cal N}=2$ supergravity without the timelike assumption.

After reviewing the ten-dimensional system of \cite{10d} in section \ref{sub:10d} (specifically in (\ref{eq:susy10})), we will discuss in sections \ref{sub:factor} and \ref{sub:flux} how to specialize it to the factorized geometry (\ref{eq:metAn}). We will then apply those considerations to each of the equations in (\ref{eq:susy10}), in sections \ref{sub:LK}, \ref{sub:psp10}, \ref{sub:pairing}; section \ref{sub:10dsum} is a brief summary.

\subsection{Ten-dimensional equations} 
\label{sub:10d}

We will begin by reviewing the system found in \cite{10d}, in a slightly different formulation.

From the ten-dimensional supersymmetry parameters $\epsilon_i$, one can define the vectors (or one-forms)
\begin{equation}
	K_i^M \equiv \frac1{32}\overline{\epsilon_i}\gamma^M \epsilon_i \ ; 
	\qquad K \equiv \frac12 (K_1 + K_2 ) \ ,\qquad
	\tilde K \equiv \frac12 (K_1 - K_2)\ .
\end{equation}
One can also define the polyform
\begin{equation}
	\Phi \equiv \epsilon_1 \overline{\epsilon_2}\ .
\end{equation}
$\Phi$ can define many different $G$-structures on the tangent bundle, but all of them correspond to the same structure on the generalized tangent bundle $T \oplus T^*$. This is similar to the application of generalized complex geometry to vacuum solutions \cite{gmpt2}.  However, similarly to what we saw in four dimensions, each of the $\epsilon_i$ is not enough to define a complete vielbein. In order to complete them and define a metric, one needs to introduce extra vectors $e^-_1$, $e^-_2$, which are null and with a fixed projection over $K_1$, $K_2$:\footnote{Notice that in \cite{10d} the vectors $e_+^i$ are used instead of $e_i^-$. The two choices are of course related by the relations $e_i^- = g^{-+} e_{+i} = 2 e_{+i}$.}
\begin{equation}\label{eq:e-10}
	(e^-_i)^2=0\ ,\qquad e^-_i \cdot K_i = 1\ ;
\end{equation}
in section \ref{sub:LK} we will obtain the identification $K_i = \frac 12 k_i$ and so we can see that (\ref{eq:e-10}) are analogous to (\ref{eq:e-}) in four dimensions.

The idea is now to reformulate the supersymmetry conditions in terms of the data $(\Phi, e^-_1, e^-_2)$. We will now give the differential equations they have to obey, in a slightly different form than in \cite{10d}:
\begin{subequations}\label{eq:susy10}
	\begin{align}
		\label{eq:LK}&L_K g = 0 \ ,\qquad d\tilde K = \iota_K H \ ;\\
		\label{eq:psp10}
		&d_H(e^{-\phi} \Phi) = -(\tilde K\wedge + \iota_K ) F \ ;\\
		& \label{eq:++1} 
		( \psi_- \otimes \overline{\epsilon_2} e^-_2\, ,\,  \pm d_H(e^{-\phi} \Phi \cdot e^-_2) + \sigma_2 \Phi - 2F)=0\ ,\\
		& \label{eq:++2}
		(e^-_1 \epsilon_1 \otimes \overline{\psi_-}\, ,\, d_H(e^{-\phi} e^-_1\cdot\Phi) - \sigma_1 \Phi - 2F )=0\ , \qquad \forall \psi_- \in \Sigma_-\ .
	\end{align}
\end{subequations}
((\ref{eq:LK}) had already been discussed in \cite{hackettjones-smith,figueroaofarrill-hackettjones-moutsopoulos,koerber-martucci-ads}.) Here, $\phi$ is the dilaton, and we have introduced the notation 
\begin{equation}
	\sigma_i \equiv \frac12 e^\phi d^\dagger (e^{-2 \phi} e^-_i)\ ;
\end{equation}
$H$ is the NSNS three-form, and $d_H\equiv d-H\wedge$. $F$ is the total RR field strength $F=\sum_k F_k$ (where the sum is from $0$ to $10$ in IIA and from 1 to 9 in IIB), which is subject to the self-duality constraint 
\begin{equation}\label{eq:F*F}
	F=*_{10} \lambda (F) \ .
\end{equation}
$(\,,\,)$ is the usual Chevalley-Mukai pairing of forms defined in (\ref{eq:mukai}). Finally, $\psi_-$ is any ten-dimensional spinor of negative chirality, and  $\psi_- \otimes \overline{\epsilon_2} e^-_2$ and $e^-_1 \epsilon_1 \otimes \overline{\psi_-}$ are then a certain set of differential forms, which were characterized a little differently in \cite{10d}.\footnote{Considering for example \cite[(3.1c)]{10d}, one can take $\gamma^{MN}$ to the left of the pairing, and then one has the set of forms $\{\gamma^{MN} e^-_1 \epsilon_1 \overline{\epsilon_2} e^-_2, \forall M,N=0,\ldots,9 \}$. But $\{\gamma^{MN} e^-_1 \epsilon_1, \forall M,N=0,\ldots,9 \}$ just coincides with the space of ten-dimensional spinors $\psi_-$ of negative chirality, since there is only one non-zero chiral orbit for ${\rm Spin}(1,9)$.}

Equations (\ref{eq:susy10}) are \emph{necessary and sufficient} for supersymmetry to hold \cite{10d}. To also solve the equations of motion, one needs to impose the Bianchi identities, which away from sources (branes and orientifolds) read 
\begin{equation}\label{eq:bianchi}
	dH=0 \ ,\qquad d_H F=0\ .
\end{equation}
It is then known (see \cite{lust-tsimpis} for IIA, \cite{gauntlett-martelli-sparks-waldram-ads5-IIB} for IIB) that almost all of the equations of motion for the metric and dilaton follow.

Working with the generalized tangent bundle $T \oplus T^*$ usually means that the natural local symmetry group gets enlarged from the Lorentz group to a ``generalized Lorentz group'' ${\rm SO}(10,10)$. The new generators include the so-called $b$-transform 
\begin{equation}\label{eq:Phib}
	\Phi \to  e^{b\wedge} \Phi\ .
\end{equation}
This is a symmetry of (\ref{eq:psp10}) if we also perform the transformations
\begin{equation}\label{eq:Hb}
	H = H - d b \ ,\qquad
	K \to K \ ,\qquad \tilde K \to \tilde K + \iota_K b \ ,\qquad
	F \to e^{b\wedge} F \ , 
\end{equation}
although now $F$ should satisfy $F=*_b \lambda F $ rather than (\ref{eq:F*F}), where $*_b= e^{b\wedge} * \lambda e^{-b\wedge} \lambda$. This is also a symmetry of (\ref{eq:LK}), provided 
\begin{equation}\label{eq:LKb}
	L_K b =0 \ .
\end{equation}
Making this symmetry work for (\ref{eq:++1}) and (\ref{eq:++2}) is more complex in general. However, we will see in section \ref{sub:flux} that it does indeed work when $b$ is purely internal.


\subsection{Factorization} 
\label{sub:factor}

As anticipated, we will consider a metric of the form (\ref{eq:metAn}), where $x^{\mu}$ are the coordinates on the four-dimensional space-time $M_4$ and $y^{m}$ are the coordinates on the internal manifold $M_6$. 

As usual, in this situation we can decompose gamma matrices as
\begin{equation}\label{eq:Gamma46}
	\Gamma_{\mu}^{(10)} = \gamma_\mu^{(4)} \otimes 1^{(6)}\ , \qquad \Gamma_{m}^{(10)} = \gamma_5^{(4)} \otimes \gamma_m^{(6)}\ .
\end{equation}
($\gamma_5^{(4)}$ was called simply $\gamma$ in section \ref{sec:4d}.) One can now decompose the ten-dimensional supersymmetry parameters $\epsilon_i$ as sums of tensor products. In general this sum would require four four-dimensional spinors for each $\epsilon_i$. However, we want to stay as close as possible to ${\cal N}=2$ supergravity, which has only two four-dimensional supersymmetry parameters $\zeta_i$. For this reason we will use the Ansatz
\begin{align}\label{eq:spAn}
\begin{split}
	&\epsilon_1 = \zeta_1 (x)\, \eta^1_+ (x,y) \,+\, \zeta^1(x)\,\eta^1_-(x,y)  \\ 
    & \epsilon_2 = \zeta_2 (x)\, \eta^2_\mp (x,y) \,+\, \zeta^2(x)\,\eta^2_\pm(x,y)\ .
\end{split}
\end{align}
Here $\zeta_i$ are spinors on $M_4$ of positive chirality (they are the same spinors the we introduced in section \ref{sec:4d}), and $\eta^i_+$ are spinors on $M_6$ of positive chirality, while $\eta^i_- = (\eta^i_+)^*$ are their Majorana conjugates, so that $\epsilon_i$ are Majorana.\footnote{We work in a basis where gamma matrices are real in four dimensions and purely imaginary in six, so that Majorana conjugation is just naive complex conjugation.} Notice that ${\cal N}=1$ flux vacua (namely, solutions where $M_4$ is a maximally symmetric space, Minkowski$_4$ or AdS$_4$), can be obtained from (\ref{eq:spAn}) by setting $\zeta_1=\zeta_2$. However, for solutions with four supercharges which are not vacua, one could use a more general Ansatz involving for example four $\zeta_i$ obeying some constraints.


The spinor Ansatz (\ref{eq:spAn}) immediately lets us compute some of the ingredients of (\ref{eq:susy10}): namely, $\Phi, K, \tilde K$. First we evaluate $\Phi= \epsilon_1 \otimes \overline{\epsilon_2}$:
\begin{equation}\label{eq:Phi46}
\begin{split}
	\Phi &= 
	2 {\rm Re} [ \mp \zeta_1 \overline{\zeta^2} \wedge \phi_\mp + (\zeta_1 \overline{\zeta_2}) \wedge \phi_\pm] \\
	 &=2 {\rm Re} \left[\mp (v+ i * v) \wedge \phi_\mp + \Big(\mu\,(1 + i\vol_4) + \omega\Big) \wedge \phi_\pm\right]\ , 
\end{split}
\end{equation}
where we have used (\ref{eq:bisp12}), and, as in \cite{gmpt2}, 
\begin{equation}
	\phi_{\pm}= \eta^1_+ \eta^{2\,\dagger}_\pm\ 
\end{equation}
are the six-dimensional pure spinors, which together define an ${\rm SU}(3) \times {\rm SU}(3)$ structure. The origin of the signs in (\ref{eq:Phi46}) is explained in appendix \ref{sec:Phi46}. 

Let us now compute $K_1$ and $K_2$. As in the case of ${\cal N}=1$ vacua \cite[Sec.~4.1.2]{10d}, the six-dimensional components of these vectors vanish ($K_i^m=0$, $m=1,\ldots,6$) because $\eta^\dagger_- \gamma^m \eta_+ = 0$. For external indices we have 
\begin{align}
	K_{1}^{\mu} = \frac{1}{4} k_1^{\mu} ||\eta^1||^2\ , \qquad K_{2}^{\mu} = \frac{1}{4} k_2^{\mu} ||\eta^2||^2\ .
\end{align}
We now assume for simplicity that the norms of the $\eta^a$ are equal:
\begin{equation}\label{eq:eqnorms}
	|| \eta^1 ||^2 = || \eta^2 ||^2\ .
\end{equation} 
We do not expect that allowing unequal norms would lead to a substantial change in our discussion. For ${\cal N}=1$ vacua, one can actually even show that (\ref{eq:eqnorms}) is necessary \cite[App.~A.3]{gmpt3}. $K$ and $\tilde{K}$ take then the form
\begin{equation}\label{eq:c}
	K^\mu = c\, \bigl(k_1^\mu + k_2^\mu \bigr)\ , \qquad 
	\tilde{K}^\mu = c\, \bigl(k_1^\mu - k_2^\mu \bigr)\ , \qquad  c \equiv \frac{||\eta||^2}{8}\ .
\end{equation}
The name $c$ anticipates that we will soon find that it has to be a constant.


\subsection{Fluxes} 
\label{sub:flux}

The next ingredient of (\ref{eq:susy10}) we need to consider is the RR polyform $F$, which is a formal sum of all fluxes of different degrees. In the following, it will be useful to decompose this polyform
\begin{equation}\label{eq:F012}
	F= F_0 + F_1 + F_2 + F_3 + F_4
\end{equation}
where $F_i$ is a \emph{polyform} with exactly $i$ external indices (and not the RR form with $i$ overall indices). In particular, from (\ref{eq:F*F}) we see that
\begin{equation}
	F_4 = * \lambda F_0 \ ,\qquad F_3 = * \lambda F_1 \ ,\qquad F_2 = * \lambda F_2 \ .
\end{equation}

An analogous decomposition is valid also for the three-form $H$
\begin{equation}
  H= H_0 + H_1 + H_2 + H_3\ 
\end{equation}
and for the $B$ field: $B= B_0 + B_1 + B_2$. Locally we have $H_0 = d_6 B_0$, $H_1= d_4 B_0 + d_6 B_1$, $H_2= d_4 B_1 + d_6 B_2$, $H_3= d_4 B_2$. 

We will now make a few assumptions about these fluxes. Since we have already assumed $\del_m g_{\mu\nu}=0$ (see footnote \ref{foot:fibr}), it is natural to also assume $\del_m B_{\mu\nu}=0$, or in other words $d_6 B_2=0$. We have also assumed that the fibration is trivial; the analogue of this is to assume that $B_1=0$. So locally we now have  
\begin{equation}\label{eq:H1H20}
	H_2= 0 \ ,\qquad H_1 = d_4 B_0 \ .
\end{equation}
In this situation, $d_H= d_4 + d_4 B_0 + d_6 + d_6 B_1 + H_3\wedge= e^{-B_0\wedge} d e^{B_0 \wedge} + H_3 \wedge$. 

This suggests that it is convenient to use the $b$-transformation (\ref{eq:Phib}), (\ref{eq:Hb}) with $b=B_0$, and work with a system where the differential is $d+ H_3 \wedge$, and $\Phi$ is replaced by $e^{B_0}\wedge \Phi$. We have already seen that this is a symmetry for (\ref{eq:LK}), (\ref{eq:psp10}), as long as (\ref{eq:LKb}), which for us reads $L_K B_0=0$, which we will show in subsection \ref{sub:LK}. For (\ref{eq:++1}), (\ref{eq:++2}), in general the $b$-transform is not a symmetry, but for $b=B_0$ (namely, purely internal $B$ field) we will see in section \ref{sub:pairing} that it is.

Hence from now on we will work with $e^{B_0}\wedge \Phi$ rather than with $\Phi$. This means that (\ref{eq:Phi46}) is now modified by modifying the internal pure spinors 
\begin{equation}\label{eq:phipmB0}
	\phi_\pm \to \phi_\pm^{B_0} \equiv e^{B_0}\wedge \phi_\pm \ .
\end{equation}
Actually, however, since from now on we will only work with the $\phi^{B_0}_\pm$, we will drop the ${}^{B_0}$ superscript and write simply $\phi_\pm$ in all our equations.


\subsection{Symmetry equations: (\ref{eq:LK})} 
\label{sub:LK}

Having specified in section \ref{sub:factor} our Ansatz (\ref{eq:metAn}), (\ref{eq:spAn}), and what it implies on the various ingredients in the supersymmetry equations (\ref{eq:susy10}), we can now start seeing what those equations become. 

We will start by (\ref{eq:LK}). The condition that $K$ is a Killing vector for the ten-dimensional metric (\ref{eq:metAn}) splits according to whether the indices are both external, both internal, or one of each. Recalling (\ref{eq:c}), the last case gives
\begin{subequations}\label{eq:Kkilling}
\begin{equation}
 	\nabla_m || \eta ||^2 k_\mu = 0 \ .\label{eq:Kkilling1}	
\end{equation}	
Thus $|| \eta ||^2$ does not depend on the internal coordinates. The residual dependence of $||\eta||^2$ from the external coordinates can be  reabsorbed in the definition of the four-dimensional spinors; hence we can simply assume it is a constant (as anticipated in (\ref{eq:c})). In the following we will fix systematically $||\eta||^2=2$ in order to have $\bar{\epsilon}_i e_i^- \epsilon_i=32$ as in \cite{10d}. Notice that, with this choice, we have $e_i^- \cdot k_i=2$ as in four dimensions, a consequence that we anticipated around (\ref{eq:e-10}). In the following we will make use of names such as $k_i$, $k$ and $v^3$ in order to stay closer to section \ref{sec:4dsugratime}, so that (\ref{eq:c}) now reads $K=\frac12 k$, $\tilde K = \frac12 v^3$. To be precise, however, these vectors are not exactly the same as those we encountered in four dimensions, because in four dimensions the natural metric would be the one in the Einstein frame, which differs from the $ds^2_4$ in (\ref{eq:metAn}) by a function of the dilaton. Since our aim is not to reduce the ten-dimensional theory to four dimensions, however, we will not perform any rescaling and will work with the string frame metric. In (\ref{eq:++1}) and (\ref{eq:++2}), we also have two extra vectors $e^-_1$, $e^-_2$, which we have assumed to satisfy (\ref{eq:e-10}). Given that the $k_i$ are purely external, (\ref{eq:e-10}) can be satisfied by simply taking the $e^-_i$ to be purely external as well, and in fact to be equal (up to the rescaling just discussed) to the vectors by the same name that we introduced in section \ref{sec:4d} (see for example (\ref{eq:e-})).

Coming back to (\ref{eq:LK}), the purely external and purely internal case give
\begin{align}
	    & \nabla_{(\mu} k_{\nu)} = (L_k g_4)_{\mu\nu} = 0 \ , \label{eq:Kkilling2}  \\
		& \nabla_{(m} k_{n)} = (L_k g_6)_{mn} = 0\ : \label{eq:Kkilling0}
\end{align}
\end{subequations}
namely, $k$ is a Killing vector for the four-dimensional metric $g_{\mu\nu}$, and it is a symmetry for the internal metric $g_{mn}$ as well.

Using our assumption (\ref{eq:H1H20}), the equation $d\tilde K = i_K H$ yields
\begin{subequations}\label{eq:dKtilde}
\begin{align}
	& d_4 \tilde k = \iota_k H_3 \label{eq:dKtilde2} \ , \\
	& 0 = \iota_k H_1 \label{eq:dKtilde0}\ .
\end{align}
\end{subequations}
(\ref{eq:dKtilde0}) can also be written as $L_k B_0=0$, which we promised at the end of section \ref{sub:flux}. We will also see later that, in the timelike case, (\ref{eq:dKtilde0}) and (\ref{eq:Kkilling0}) both follow from other equations (namely, from invariance under $k$ of the internal pure spinors). 


\subsection{Exterior equation: (\ref{eq:psp10})} 
\label{sub:psp10}

From now on, in this section and in the next, for simplicity of notation we will specialize to IIA; the IIB case is very similar, and differs by some signs only.

In a similar manner as we did for equations (\ref{eq:LK}), equation (\ref{eq:psp10}) splits in five pieces according to the number of external components involved:
\begin{subequations}\label{eq:ext}
\begin{align}
	\label{eq:ext0}
	& d_6 {\rm Re} \bigl( e^{-\phi}\mu \phi_+ \bigr ) = - \frac14 i_k F_1 \ , \\
	\label{eq:ext1}
	&  d_4 {\rm Re} \bigl( e^{-\phi}\mu \phi_+ \bigr) - d_6 {\rm Re}  \bigl(e^{-\phi} v \wedge \phi_- \bigr) = - \frac14 \bigl(i_{k} F_2  + v^3 \wedge F_0 \bigr) \ , \\
	\label{eq:ext2}
 	& - d_4 {\rm Re}  \bigl(e^{-\phi} v \wedge \phi_-\bigr) + d_6 {\rm Re}  \bigl(e^{-\phi} \omega \wedge \phi_+ \bigr) = - \frac14 \bigl(i_k F_3 + v^3 \wedge F_1 \bigr) \ , \\
	 \label{eq:ext3}
	\begin{split}
		&  d_4 {\rm Re} \bigl(e^{-\phi} \omega \wedge\phi_+ \bigr) - d_6 {\rm Re}\bigl( e^{-\phi}i * v \wedge \phi_- \bigr)  \\
		  & \hspace{3cm}+ H_3 \wedge {\rm Re}  \bigl(e^{-\phi} \mu \wedge \phi_+ \bigr) = - \frac14 \bigl( i_k F_{4} + v^3 \wedge F_2\bigr)\ ,
	\end{split}\\
	\label{eq:ext4}
	\begin{split}
		&  - d_4 {\rm Re} \bigl(e^{-\phi} i * v \wedge \phi_- \bigr) + d_6 {\rm Re} \bigl( e^{-\phi} i \mu \vol_4 \wedge \phi_+ \bigr)  \\
	  & \hspace{3cm}- H_3 \wedge {\rm Re} \bigl(e^{-\phi}v \wedge\phi_- \bigr)= -\frac14 v^3 \wedge F_3\ .
	\end{split}
\end{align}
\end{subequations}
Recall that the $\phi_\pm$ here (and in the equations that will follow in the rest of the paper) include the internal $B_0$ field as in (\ref{eq:phipmB0}).

For the particular case of four-dimensional vacua, where the dependence on $M_4$ is trivial, (\ref{eq:ext}) reduce to the pure spinor equations of \cite{gmpt2}. In general, they contain information both about the geometry of $M_4$ (via $d_4$ of the external forms) and about the way the metric of $M_6$ changes as a function of the spacetime coordinates $x^\mu$ (via $d_4 \phi_\pm$). In particular, in the timelike case we will see that (\ref{eq:ext1}) and (\ref{eq:ext2}) give first-order equations for the dependence of the scalars in the vector multiplets and hypermultiplets respectively. These first-order equations would give rise to attractor-like equations in black hole applications.


\subsection{Pairing equations: (\ref{eq:++1}), (\ref{eq:++2})} 
\label{sub:pairing}

We now turn to (\ref{eq:++1}), (\ref{eq:++2}). As noted in section \ref{sub:LK}, we can take $e^-_i$ to be purely four-dimensional. In this subsection, we will put $\phi = H_3 =0$ for simplicity. This simplification can be made with no loss of information for the following reason: equations (\ref{eq:pairingf1}), (\ref{eq:pairingf2}), (\ref{eq:pairingf3}) and (\ref{eq:pairingf4}), which we will further analyse in section \ref{sec:time10d}, are not modified by this simplification. On the other hand equations (\ref{eq:++1q}), (\ref{eq:++1p}), (\ref{eq:++2q}) and (\ref{eq:++2p}) will be useful for us only to suggest a system of equations for the general case in four dimensions (which is the subject of section \ref{sec:4dgen}). Therefore the details of these equations are not important for our purposes.

We will start by considering (\ref{eq:++1}). We should give a convenient basis of possible ten-dimensional Majorana--Weyl spinors $\psi_-$ of negative chirality. Given our spinor decomposition (\ref{eq:spAn}), we can give this list as
\begin{equation}\label{eq:reimchi}
	\psi_- = \left\{ \begin{array}{c}
		\chi_- + \chi_-^* \ ,\\
		i(\chi_- - \chi_-^*)\ ,
	\end{array}	\right.
\end{equation}
where $\chi_-$ should now run over a basis of negative chirality spinors. To obtain such a basis, we can tensor a four-dimensional basis with a six-dimensional one. Given a Weyl spinor $\zeta_+$ of positive chirality, in the notation of section \ref{sub:4deps},
\begin{equation}
\label{eq:+basis}
\zeta_+\ , \qquad e^ - \cdot  \zeta_-
\end{equation}
is a basis for the space $\Sigma_+$ of spinors of positive chirality, and 
\begin{equation}
\label{eq:-basis}
\zeta_-\ , \qquad e^- \cdot  \zeta_+ \, 
\end{equation} 
is a basis for the space $\Sigma_-$ of spinors of negative chirality. In six dimensions, on the other hand, we have the bases
\begin{equation}\label{eq:6dbases}
	\{ \eta^1_+ , \gamma^m \eta^1_-\} \ ,\qquad \{ \eta^1_- , \gamma^m \eta^1_+\} 
\end{equation}
for spinors of positive and negative chirality respectively. Actually, in (\ref{eq:6dbases}) only three out of six $\gamma^m$ give a non-zero result: $\gamma^m \eta^1_+$ is non-zero only if the index $m$ is anti-holomorphic with respect to the almost complex structure defined by $\eta^1_+$. So it would be more precise to write $\gamma^{\bar{i}_1} \eta^1_+$, as in \cite[App.~A.4]{gmpt3}.

Summing up, in (\ref{eq:reimchi}), $\chi_-$ can be any of the following: 
\begin{equation}\label{eq:46basis}
	\zeta_1 \eta^1_- \ ,\qquad \zeta_1 \gamma^m \eta^1_+ \ ,\qquad
	e^-_1 \zeta^1 \eta^1_- \ ,\qquad e^-_1 \zeta^1\gamma^m \eta^1_+\ .
\end{equation}

As a further simplification, notice that, when we plug (\ref{eq:reimchi}) in (\ref{eq:++1}), the two equations $((\chi_- + \chi_-^*) \overline{\epsilon_2}e^-_2,\ldots)$ and $i((\chi_- - \chi_-^*) \overline{\epsilon_2}e^-_2, \ldots)$ are simply the real and imaginary parts of 
\begin{equation}\label{eq:chi++}
	(\chi_- \bar{\epsilon}_2 e_2^-, d_H(e^{-\phi} \Phi \cdot e^-_2) + \sigma_2 \Phi - 2 F ) = 0\ .
\end{equation}

So all we have to do is to impose (\ref{eq:chi++}) for all $\chi_-$ in (\ref{eq:46basis}). We will start with 
\begin{equation}
	\chi_- = \zeta_1 \eta^1_-\ .
\end{equation}
In this case we compute (in a similar way as in appendix \ref{sec:Phi46}):
\begin{equation}\label{eq:firstchi}
	\chi_- \overline{\epsilon_2} e^-_2 = 
	\zeta_1 \overline{\zeta^2} e^-_2 \wedge \bar \phi_+ + 
	\zeta_1 \overline{\zeta_2} e^-_2 \wedge \bar \phi_- \ .
\end{equation}
At this point, these are the old $\phi_\pm$, before implementing the $B_0$ transform (\ref{eq:phipmB0}) that we advocated at the end of section \ref{sub:flux}. However, it is now easy to show that (\ref{eq:phipmB0}) is a symmetry of the pairing equations. The term $d_H(e^{-\phi} \Phi \cdot e^-_2) + \sigma_2 \Phi$ in the right entry of (\ref{eq:chi++}) simply becomes $e^{-B_0}\wedge (d_H(e^{-\phi} e^{B_0}\wedge\Phi \cdot e^-_2) + \sigma_2 \Phi)$. The $e^{-B_0}$ can be sent to the left entry, where it becomes a $e^{B_0}$. We now have to also multiply $F$ in the right entry by $e^{B_0}$, if we remember that it should now satisfy $e^{-B_0} F= * \lambda e^{-B_0}F$ rather than (\ref{eq:F*F}). All the $e^{B_0}$ factors can now be reabsorbed in the $\phi_\pm$ as in (\ref{eq:phipmB0}). As we preliminarily declared after (\ref{eq:phipmB0}), from now on our $\phi_\pm$ will actually be understood as $e^{B_0}\wedge \phi_\pm$.

We can now evaluate (\ref{eq:chi++}) with (\ref{eq:firstchi}). Let us define
\begin{equation}\label{eq:SfT}
	S_3 \equiv i (\bar\phi_+ , F_0) \ ,\qquad f_1 = dx^\mu (\bar \phi_- , F_{1\,\mu})\ ,\qquad
	T^+_{\mu\nu} \equiv i (\bar \phi_+ , F_{2\,\mu\nu}) \ .
\end{equation}
These definitions here appear only as a notational simplification but they will be motivated in section \ref{sub:org}. In this manner we obtain\footnote{Recall the following relations (see for example \cite[Eq.~(3.41)]{gmpt3})
\begin{equation}\label{eq:normpurespinor}
  (\phi_\pm,\bar{\phi}_\pm) = -\frac{i}{8}||\eta^1_\pm||^2||\eta^2_\pm||^2 = -\frac i2
\end{equation}
and the last equality makes use of $||\eta||^2=2$.
}
\begin{subequations}\label{eq:++1tot}
\begin{equation}\label{eq:++1q}
	e^-_2 \cdot (-8 q_1 + S_3 v + \iota_v T^+) + i (\mu e^-_2 - \iota_{e^-_2}\omega) \cdot f_1=0\ ,
\end{equation}
where $e^-_2 \cdot q_1$ can be computed in terms of derivatives of forms using (\ref{eq:eq}).

In a similar way, the other spinors in (\ref{eq:46basis}) give the equations 
\begin{align}
	\label{eq:++1p}
	&8 e^-_2 \cdot \Big(  p_1 -  i (\bar \phi_+, d_4 \phi_+) -  i (\bar \phi_-, d_4 \phi_-)\Big) - \iota_{e_1^-}(\bar{\mu} e_2^- - \iota_{e_2^-} \bar{\omega}) \llcorner T^+  = -\bar \nu S_3  -i  (2 e^-_1 \cdot \bar v e^-_2 -\bar{\nu} w_2 )\cdot f_1\ ,\\ 
   \label{eq:pairingf1}& e^-_2 \cdot \Big( v (\gamma^{\bar{i}_1}\phi_-, F_0 ) + \iota_v (\gamma^{\bar{i}_1}\phi_- , F_2) \Big) + (\mu e^-_2 - \iota_{e^-_2} \omega) \cdot (\gamma^{\bar{i}_1}\phi_+ , F_1)=0\ ,\\
   \label{eq:pairingf2}& -\bar \nu (\gamma^{\bar{i}_1}\phi_-, F_0) + \iota_{e^-_1}(\bar \mu e^-_2 - \iota_{e^-_2} \bar \omega) \llcorner (\gamma^{\bar{i}_1}\phi_-, F_2) = (2 e^-_1 \cdot \bar v e^-_2 - \bar{\nu}  w_2 )\cdot (\gamma^{\bar{i}_1}\phi_+ , F_1) \ ,
\end{align}
\end{subequations} 
where 
\begin{equation}
	\nu \equiv  \mu e^-_1 \cdot e^-_2 - \iota_{e^-_1}\iota_{e^-_2} \omega \ .
\end{equation}
Together, (\ref{eq:++1q}) are all the components of (\ref{eq:++1}) relevant to the Ansatz (\ref{eq:spAn}) introduced in this section. 

One can deal with (\ref{eq:++2}) in a similar way. We get  
\begin{subequations}\label{eq:++2tot}
\begin{align}
		\label{eq:++2q}
		&e^-_1 \cdot ( -8 \bar q_2 + \bar{S}_3 v + \iota_v T^- ) + i (\bar \mu e^-_1 + \iota_{e^-_1} \bar \omega) \cdot f_1 = 0 \ , \\
		\label{eq:++2p}
	   & e^-_1 \cdot \Big( - 8 \bar p_2 + 8 i (\phi_+, d_4 \bar\phi_+) + 8 i (\bar \phi_-, d_4 \phi_-) - (\mu e_2^- - \iota_{e^-_2} \omega) \llcorner T^- \Big) = - \nu \bar S_3 - i  
(2 e^-_2 \cdot \bar v e^-_1 + \nu \bar w_1 )\cdot f_1\ , \\
	\label{eq:pairingf3}& e^-_1 \cdot \Big( v (\phi_- \gamma^{\bar j_2}, F_0) - \iota_v (\phi_- \gamma^{\bar j_2}, F_2)\Big) + ( \bar \mu e^-_1 + \iota_{e^-_1} \bar \omega) \cdot (\bar\phi_+ \gamma^{\bar j_2},F_1)=0\ ,\\
	 \label{eq:pairingf4}& \nu (\phi_- \gamma^{\bar j_2}, F_0) - \iota_{e^-_1} (\mu e^-_2 - \iota_{e^-_2} \omega) \cdot (\phi_- \gamma^{\bar j_2}, F_2) = -(2 e^-_2 \cdot \bar v e^-_1 + \nu \bar w_1 )\cdot (\bar \phi_+ \gamma^{\bar j_2}, F_1) \ , 
\end{align}	
\end{subequations}
where $T^- \equiv \overline{ T^+}$.



\subsection{Summary: ten-dimensional system} 
\label{sub:10dsum}

In this section, we have applied the ten-dimensional system \cite{10d} to a (topologically trivial) fibration (\ref{eq:metAn}), with a spinor Ansatz (\ref{eq:spAn}), (\ref{eq:eqnorms}), and a few assumptions summarized in footnote \ref{foot:fibr} and in (\ref{eq:H1H20}). The conditions of preserved supersymmetry are equivalent to equations (\ref{eq:Kkilling}), (\ref{eq:dKtilde}), (\ref{eq:ext}), (\ref{eq:++1tot}), (\ref{eq:++2tot}). (These last two were given with the additional assumption $\phi=H_3=0$.) These equations are not as pleasant as one might wish, but fortunately we will be able to do much better in the next section. There, we will apply the system in the timelike case, and massage it to a much more pleasant-looking form, which will closely parallel the ``boxed'' system seen in section \ref{sec:4dsugratime}. 



\section{Ten-dimensional system in the timelike case} 
\label{sec:time10d}

In this section, we will rewrite the equations we found in section \ref{sec:10d} in the timelike case. The conditions for supersymmetry are expected to be much simpler in this case. Most of the equations we will obtain organize themselves in a way that closely parallels the boxed system (\ref{eq:timesusy}), (\ref{eq:timesusyhyper}), (\ref{eq:timesusygaugino}) in section \ref{sec:4dsugratime}. As we will see, however, there will also be equations formally associated with ``gravitino multiplets''. 

We will start in section \ref{sub:org} with a discussion of how ten-dimensional fields organize themselves on a spacetime $M_4 \times M_6$, where $M_6$ is a ${\rm SU}(3) \times {\rm SU}(3)$ structure manifold. Many fields come from forms: the RR fields, but also the internal metric and $B$ field, through the pure spinors $\phi_\pm$. A useful basis for internal forms is given by the ``generalized Hodge diamond'', (\ref{eq:hodge}) below. The most substantial part of the multiplets will correspond to the interior of that diamond. The edges are usually discarded in ${\cal N}=2$ reductions for reasons we will review below; however, in this paper we are not performing a reduction, and we will need to keep the corresponding representations from the edges.

Section \ref{sub:org} will then dictate the way we organize our ten-dimensional equations in later subsections. We will describe those corresponding to the four-dimensional gravitino equations in section \ref{sub:4dgeom}, and to the universal and non-universal hypermultiplets in sections \ref{sub:univ} and \ref{sub:otherhyp}. We will then have vector multiplets from the bulk of the diamond (section \ref{sub:vm}) and our new vector multiplets from the edge (section \ref{sub:edge-vm}). Finally, we will have in section \ref{sub:edge-gino} some new equations associated with gravitino multiplets, in a sense we will clarify.

Section \ref{sub:fibr} will describe how to extend our results to nontrivial fibrations. We will then summarize our results in section \ref{sub:10dtimelikesummary}.

\subsection{Organizing the fields} 
\label{sub:org}

We will first review how the ten-dimensional fields produce the various four-dimensional fields in a reduction, and then how these get organized in multiplets for ${\cal N}=2$ compactifications. Most of this material is by now standard. One purpose in reviewing it here is to introduce a few definitions that will be useful later. Another purpose is that some of our equations will be in ``vector'' representations associated to the edge of the diamond in (\ref{eq:hodge}) below; these will organize themselves in multiplets which are not commonly considered in the literature, as we will see. 

\subsubsection{Scalars} 
\label{ssub:scalars}

We will start by considering spacetime scalars that come from deforming the internal NSNS fields, $g_{mn}$ and $B_{0\,mn}$. These degrees of freedom are determined \cite{gualtieri} (see \cite[Sec.~3]{gmpt3} for a review) by the internal pure spinors $\phi_\pm$, along with the internal dilaton and spinors: 
\begin{equation}
	\{ g_{mn}, B_{0\,mn}, \phi, \eta^{1,2}\} \leftrightarrow \phi_\pm
\end{equation}
Hence the deformations $\delta g_{mn}$, $\delta B_{0\,mn}$ come from the deformations $\delta \phi_\pm$ of the pure spinors. We hence need to expand these latter deformations in an appropriate basis for internal forms.

The most natural basis is given by the so-called generalized Hodge diamond:
\begin{equation}
  \label{eq:hodge}
  \begin{array}{c}\vspace{.1cm}
\phi_+ \\ \vspace{.1cm}
\phi_+\gamma^{i_2}  \hspace{1cm} \gamma^{\bar i_1}\phi_+ \\
\phi_-\gamma^{\bar i_2} \hspace{1cm} \gamma^{\bar i_1} \phi_+\gamma^{i_2}
\hspace{1cm} \gamma^{i_1} \bar \phi_-\\
\phi_- \hspace{1.2cm}\gamma^{\bar i_1}\phi_-\gamma^{\bar j_2}
\hspace{1cm} \gamma^{i_1}\bar\phi_-\gamma^{j_2}
\hspace{1.2cm}\bar\phi_-\\
 \gamma^{\bar i_1} \phi_-\hspace{1cm} \gamma^{i_1} \bar \phi_+ \gamma^{\bar j_2}
\hspace{1cm}\bar\phi_-\gamma^{i_2}\\
\gamma^{i_1} \bar\phi_+ \hspace{1cm}  \bar\phi_+\gamma^{\bar i_2}\\
 \bar\phi_+\\
  \end{array}\
\end{equation}
This basis has the property that it is orthogonal: every form has vanishing six-dimensional pairing with every form in the diamond, except with the ones symmetric with respect to the central point. So for example $\phi_+$ has non-vanishing pairing only with $\bar{\phi}_+$, $\phi_+\gamma^{i_2}$ has non-vanishing pairing with $\bar{\phi}_+ \gamma^{\bar{i}_2}$ and so on.

It should be emphasized that the basis (\ref{eq:hodge}) is a basis at every point: if we expand a form in $M_6$ in terms of (\ref{eq:hodge}), we will get functions, not numbers. Nevertheless, the basis provides a way to neatly organize our equations. As mentioned in the introduction, this is similar in spirit to the pre-truncation computations in \cite{grana-louis-waldram} (corresponding to their section 2). 

We can now expand $\delta \phi_\pm$ in the basis (\ref{eq:hodge}). It should be noticed, however, that each variation can only produce forms that are ``not too far'' from the original pure spinor. Namely, $\delta \phi_+$ can only contain spinors which can be written as $\Gamma_{MN} \phi_+$, where $\Gamma_M$ are the generators of Cl$(6,6)$. Concretely, this means that it can only contain forms in the zeroth and second row of (\ref{eq:hodge}). In the same way, $\delta \phi_-$ can only contain forms in the zeroth and second column. To shorten our notation, let us introduce indices counting the (infinitely many) forms in these entries of (\ref{eq:hodge}). First of all,
\begin{equation}\label{eq:dpal+}
	\delta \phi^a_+ = \{\gamma^{i_1} \bar{\phi}_+ \gamma^{\bar{j}_2} \} \ .
\end{equation}
Let us then define\footnote{In this section the pairing $(\,,)$ denotes the six-dimensional one; we also define $(a_6, \beta_4 \wedge b_6) = \beta_4 \wedge (a_6, b_6)$, where $a_6, b_6$ are internal forms and $\beta_4$ is an external form.}
\begin{equation}\label{eq:Dt}
	 - D \bar{t}^a =  (\delta \phi^a_+, d_4 \phi_+) \qquad 	(\mathbf{3},\mathbf{\bar 3})\ .
\end{equation}
(Remember again that $\phi_+$ here actually includes a $e^{B_0}$, as in (\ref{eq:phipmB0}); hence these $t^a$ are complex. The way these scalars are defined is reminiscent of how the vector multiplet scalars in a Calabi--Yau compactification are integrals over two-cycles of the form $B_0 + i J$.)  
We have stressed the ${\rm SU}(3) \times {\rm SU}(3)$ representation in which these scalars transform. These $D\bar{t}^a$ can be morally thought of as suitable covariant derivatives of scalars $\bar{t}^a$ defined by expanding the variation of $\phi_+$ along the forms $\delta \phi^a_+$, with a connection piece coming from the fact that the $\delta \phi^a_+$ are themselves not closed. This issue potentially comes up even in Calabi--Yau compactifications, where one expands along harmonic forms $\omega_i$, which a priori should vary when one varies the metric. However, in that case the connection is flat and can be gauged away; one would want this to happen for a more general reduction to ${\cal N}=2$ supergravity \cite{kashanipoor-minasian}. Since we are not trying to reduce to an ${\cal N}=2$ effective theory, but merely to reorganize the supersymmetry equations in ten dimensions in a way inspired from ${\cal N}=2$ in four dimensions, we can afford to leave this issue unresolved. The definition (\ref{eq:Dt}) should be thought of as a bookkeeping device more than a detailed attempt at writing a four-dimensional effective theory. Similar considerations will apply to the symbols $D$ we will introduce from now on. For example, one can similarly define
\begin{equation}\label{eq:dpal-}
	\delta \phi^\alpha_- = \{ \gamma^{i_1} \bar{\phi}_- \gamma^{j_2} \} 
\end{equation}
and
\begin{equation}\label{eq:Dz}
	D (z^\alpha + i \tilde{z}^\alpha) = \mathrm{Re}(\delta \phi^\alpha_-, d_4 \phi_-) + i \mathrm{Im} (\delta \phi^\alpha_-, d_4 \phi_-)\qquad 	(\mathbf{3},\mathbf{3})\ .
\end{equation}

The expansion of $d_4 \phi_\pm$ along the forms on the boundary of the diamond (\ref{eq:hodge}) (namely, $\phi_+ \gamma^{i_2}$ and the others with one gamma acting on $\phi_\pm$) does not directly correspond to deformations of $g_{mn}$ and $B_{mn}$, but rather to changes in the spinors $\eta^{1,2}$ determined by $\phi_\pm$ (see for example \cite[Sec.~2.3]{10d}).

Finally, the expansion of $d_4\phi^a_\pm$ along the corners of (\ref{eq:hodge}), $(\bar \phi_+, d_4 \phi_+)$, will appear as a connection in some of our equations. 

Other scalars come from the RR sector. As we did earlier, it is convenient to consider the decomposition (\ref{eq:F012}) of $F$ as $\sum F_i$, where $F_i$ has $i$ external indices. From $F_1$ we have 
\begin{align}
	\label{eq:Dxia}
	D(\xi^\alpha  + i \tilde \xi^\alpha) \equiv - \frac 12 e^\phi (\delta \phi^\alpha_-, F_1) \qquad 	(\mathbf{3},\mathbf{3})\ ; \\
	\label{eq:Dxi}
	D(\xi + i \tilde \xi) \equiv  2 e^\phi (\bar \phi_-, F_1) \qquad 	(\mathbf{1},\mathbf{1})\ .
\end{align}
Notice that, for the scalars (\ref{eq:Dxia}), (\ref{eq:Dxi}), the symbol $D$ is now hiding something more than the (perhaps flat) connection we mentioned for (\ref{eq:Dt}). Here, on top of the fact that $d_4$ can act on the forms $\delta \phi^\alpha_-$, $\bar \phi_-$, we also have the fact that (locally) $F_1=d_4 C_0 + d_6 C_1$. If we defined the scalar $(\xi + i \tilde \xi)$ by $2 e^\phi (\bar \phi_-, C_0)$, we would see that (\ref{eq:Dxi}) contains a term proportional to $C_1$, which signals that the scalar is gauged under one of the spacetime vectors originated by RR fields, which we will see in section \ref{ssub:vectors}, which is in line with expectations from actual reductions in presence of internal flux. Again, since in this paper we are not actually performing a reduction, we will be content with the definition (\ref{eq:Dxi}) and will not try to resolve the symbol $D$ down to its constituents.

We also have $D$ of scalars in ``vector'' representations:
\begin{equation}\label{eq:sc31}
	(\phi_+ \gamma^{i_2}, F_1)  \qquad 	(\mathbf{1},\mathbf{3})\ ; \qquad
	(\gamma^{\bar i_1}\phi_+ , F_1)  \qquad 	(\mathbf{\bar 3},\mathbf{1})\ .
\end{equation}
These shall remain nameless, for reasons to become clear later.
Finally we have the dilaton $\phi$, and a scalar $a$ which can be defined by dualizing the spacetime NSNS three-form:
\begin{equation}\label{eq:da}
	H_3 = * d_4 a  \ .
\end{equation}
Actually this dualization procedure is only possible when $d_4 * H_3=0$. This is not guaranteed in general, since the equation of motion for $H$ reads in general $d(e^{4A-2 \phi} *H)= -e^{4A}\sum_n F_n \wedge * F_{n+2}$, and the right hand side might sometimes not vanish. This corresponds roughly to a case where one wants to include both magnetic and electric gaugings at the same time. When $H_3$ cannot be dualized, one cannot define the scalar $a$, and one would have to work with multiplets involving tensors. Although supersymmetric actions for such multiplets have been studied (see for example \cite{dallagata-dauria-sommovigo-vaula}), we will gloss over this subtlety in this paper, and assume (\ref{eq:da}).


\subsubsection{Vectors} 
\label{ssub:vectors}

The NSNS sector gives rise to four-dimensional vectors via the mixed components $g_{\mu m}$, $B_{\mu m}$. Notice that these components will be set to zero when we give our equations; however, for the time being we find it useful to consider them. 

The vectors $g_{\mu m}$, $B_{\mu m}$ both have a single internal vector index. This does not make their ${\rm SU}(3) \times {\rm SU}(3)$ representation manifest. However, writing them as $E_{\mu m}= g_{\mu m} + B_{\mu m}$, $E_{m \mu}= g_{m \mu} + B_{m \mu}=g_{\mu m} - B_{\mu m}$ and remembering the stringy origins of these fields make one guess \cite{grana-louis-waldram2} that they belong to the representations
\begin{equation}\label{eq:Emix}
	g_{\mu m} \ ,\qquad B_{\mu m}\ :  \qquad (\mathbf{3},\mathbf{1}) \oplus 
	(\mathbf{\bar 3},\mathbf{1}) \oplus (\mathbf{1},\mathbf{3}) \oplus (\mathbf{1},\mathbf{\bar 3})\ .
\end{equation}
A way to confirm this conclusion is to study explicitly how $E$ transforms under internal ${\rm O}(6,6)$ transformations $O=\left(\begin{smallmatrix}
a & b \\ c & d
\end{smallmatrix}\right)$: one obtains
\begin{equation}\label{eq:Emixtr}
	E_{\mu m} \to E_{\mu n}((c E + d)^{-1})^n{}_m \ ,\qquad E_{m \mu}\to (a - (a E + b )(c E + d)^{-1} c)_m{}^n E_{n \mu}
\end{equation}
 (where $E$ is the internal $g+B$). Using the expression for the generalized almost complex structures
\begin{equation}
	{\cal J}_{\pm} = {\cal E} \left(\begin{array}{cc}
		I_1 & 0 \\ 0 & I_2
	\end{array}\right) {\cal E}^{-1}\ ,\qquad {\cal E}= \left( \begin{array}{cc}
		1 & 1 \\ E & - E^t
	\end{array}\right)\ 
\end{equation}
(where $I_i$ are two almost complex structures), the ${\rm SU}(3) \times {\rm SU}(3)$ subgroup of ${\rm O}(6,6)$ can be characterized as 
\begin{equation}\label{eq:expU33}
	O= ({\cal E}^{-1})^t \left( \begin{array}{cc}
		U_1 & 0 \\ 0 & U_2
	\end{array}\right) {\cal E}^t\,
\end{equation}
where $U_i$ satisfy $[U_i,I_i]=0$ and $U_i^t g U_i = g$ --- namely, they are unitary with respect to the internal metric $g$. Specializing (\ref{eq:Emixtr}) to this particular $O$ leads to
\begin{equation}
	E_{\mu m} \to (U_1^{-1})_m{}^n E_{\mu n}  \ ,\qquad
	E_{m \mu} \to (U_2)_m{}^n E_{n \mu}\ ,
\end{equation}
which confirms (\ref{eq:Emix}).

We also have vectors from the RR sector.  The expansion of $F_2$ (recall that the ${}_2$ denotes the number of four-dimensional indices) gives the field-strengths 
\begin{align}
	\label{eq:t+}
	T^+ \equiv - \frac i2 e^\phi\,( \bar \phi_+ , F_2 )\ \qquad 
	(\mathbf{1},\mathbf{1})\\
	\label{eq:G+}
	G^{a-} \equiv  - \frac{i}{4}  e^\phi \, (\delta \phi^a_+, F_2) \ \qquad (\mathbf{3},\mathbf{\bar 3})
\end{align}
as well as 
\begin{equation}\label{eq:F231}
	(\phi_- \gamma^{\bar i_2}, F_2)  \qquad 	(\mathbf{1},\mathbf{\bar 3})\ ; \qquad
	(\gamma^{i_1}\bar\phi_- , F_2)  \qquad 	(\mathbf{3},\mathbf{1})\ .
\end{equation}
Similarly to our comments about (\ref{eq:Dt}), all these field strengths are not simply the exterior derivative of a potential, because of the non-constancy of $\bar \phi_+$ and $\delta \phi^a_+$, and because $F_2$ has two terms: locally, $F_2=d_6 C_2 + d_4 C_1$ (see our comment after (\ref{eq:Dxi})). As the notation implies, these are actually the self-dual (or anti-self dual) parts of the field-strengths: for example, $T^+$ satisfies $*T^+=i T^+$.


\subsubsection{Fermions} 
\label{ssub:fermions}

We also take a quick look at fermions. The spin $3/2$ fields in four dimensions originate from the ten-dimensional gravitinos, with their index taken along the four dimensions. To understand how these transform under ${\rm SU}(3) \times {\rm SU}(3)$, recall that the two SU(3) factors come from the stabilizers of the two supersymmetry parameters $\eta^1$ and $\eta^2$ respectively. That suggests that the $\psi^1$ transforms under the first SU(3) and is a singlet under the second, and that $\psi^2$ transforms under the second SU(3) and is a singlet under the first. Taking also four-dimensional chirality into account we get
\begin{equation}\label{eq:psimu}
	\begin{split}
			\psi^1_{+\mu} \qquad (\mathbf{1}, \mathbf{1}) \oplus (\mathbf{3}, \mathbf{1}) \ ; \\
			\psi^2_{+\mu} \qquad (\mathbf{1}, \mathbf{1}) \oplus (\mathbf{1}, \mathbf{\bar 3}) \ . 
	\end{split}
\end{equation}

Spin $1/2$ fields arise both from $\psi^{1,2}_m$ (the internal components of the gravitinos) and from the dilatinos $\lambda^{1,2}$. The latter transform as in (\ref{eq:psimu}):
\begin{equation}\label{eq:lambdarep}
	\begin{split}
			\lambda^1_+ \qquad (\mathbf{1}, \mathbf{1}) \oplus (\mathbf{3}, \mathbf{1}) \ ; \\
			\lambda^2_+ \qquad (\mathbf{1}, \mathbf{1}) \oplus (\mathbf{1}, \mathbf{\bar 3}) \ . 
	\end{split}
\end{equation}
The $\psi^{1,2}_m$ are subtler because we also have to work out the transformation law under ${\rm SU}(3) \times {\rm SU}(3)$ of the internal index $m$ (much as we had to do for $g_{m\mu}$ and $B_{m\mu}$ in section \ref{ssub:vectors}). As it was noticed in \cite[Sec.~5.1]{coimbra-stricklandconstable-waldram}, the correct transformation law is obtained by assuming that for $\psi^1_m$ the spinorial index transforms under the first SU(3), while the ${}_m$ index transforms under the \emph{second} SU(3); and likewise for $\psi^2_m$:
\begin{equation}\label{eq:psimrep}
	\begin{split}
			\psi^1_{+m} \qquad ((\mathbf{1}, \mathbf{1}) \oplus (\mathbf{3}, \mathbf{1})) \otimes ((\mathbf{1}, \mathbf{3}) \oplus (\mathbf{1}, \mathbf{\bar 3})) \ ; \\
			\psi^2_{+m} \qquad ((\mathbf{1}, \mathbf{1}) \oplus (\mathbf{1}, \mathbf{\bar 3})) \otimes ((\mathbf{3}, \mathbf{1}) \oplus (\mathbf{\bar 3}, \mathbf{1})) \ . 
	\end{split}
\end{equation}
This can be determined by using the ${\rm O}(d,d)$ transformation laws for fermions, and specializing them to ${\rm SU}(3) \times {\rm SU}(3)$ as in (\ref{eq:expU33}). We will not do so explicitly here, but see for example \cite[Sec.~3]{hassan-RR}.


\subsubsection{Multiplets} 
\label{ssub:multiplets}

We will now collect the vectors and scalars in four-dimensional ${\cal N}=2$ multiplets. We will not deal with the fermions, since there are non-trivial mixings between gravitinos and dilatinos \cite{grana-louis-waldram2}. 

Most multiplets are natural extensions of the ones which are familiar from Calabi--Yau compactifications. There is a vector multiplet transforming in the $(\mathbf{3}, \mathbf{\bar 3})$, which collects the scalars from (\ref{eq:Dt}), the vectors from (\ref{eq:G+}), and part of the spinors in (\ref{eq:psimrep}). There is a hypermultiplet in the $(\mathbf{3}, \mathbf{3})$, which collects the scalars in (\ref{eq:Dz}), (\ref{eq:Dxia}), and again part of the spinors in (\ref{eq:psimrep}). Finally, there is a ``universal'' hypermultiplet in the $(\mathbf{1}, \mathbf{1})$, whose scalars are (\ref{eq:Dxi}), the dilaton $\phi$, and the axion $a$ defined as usual by (\ref{eq:da}).  In the Calabi--Yau case, these would result in the usual $h^{1,1}$ vector multiplets and $1+ h^{2,1}$ hypermultiplets.

All this is standard; these multiplets were included in \cite{grana-louis-waldram2}. The situation is a bit more problematic in the ``vector'' representations, $(\mathbf{1}, \mathbf{3})$, $(\mathbf{\bar 3}, \mathbf{1})$ and their complex conjugates. These have not been included in  reductions to ${\cal N}=2$ supergravity --- for good reasons, as we will now see. Looking at sections \ref{ssub:scalars} and \ref{ssub:vectors}, the first thing we notice is that we have more vectors than could be possibly accommodated in vector multiplets: the scalars (\ref{eq:sc31}) will sit in a vector multiplet in the same representation, but their partners could be among (\ref{eq:Emix}) or perhaps (\ref{eq:F231}). 

The reason of this apparent mismatch becomes clear if we consider the case $M_6=T^6$. This produces an ${\cal N}=8$ theory. If we decompose its field content in ${\cal N}=2$ multiplets, we find 15 vector multiplets, 10 hypermultiplets, and 6 ``gravitino multiplets'' which contain a spin 3/2 field, two vectors and a spin 1/2 field. This suggests that we should include a gravitino multiplet in the $(\mathbf{3}, \mathbf{1}) \oplus (\mathbf{1}, \mathbf{3})$. (At this point we are not actually able to tell the difference between a $\mathbf{3}$ and $\mathbf{\bar 3}$, which are complex conjugates of each other.)

This does not mean we are advocating including gravitino multiplets in ${\cal N}=2$ effective theories. These multiplets are allowed classically by supersymmetry, but in general they run into trouble quantum mechanically: their spin 3/2 fields contain zero-norm states that need to be gauged out by a spinorial gauge transformation. These gauge transformations are the supersymmetry parameters; so these multiplets are only allowed when supersymmetry is actually higher, ${\cal N}>2$. Even massless gravitinos will probably arise in the context of a Higgs effect, where the spin $1/2$ gauge transformations are still present. Thus, they were not included in \cite{grana-louis-waldram2} for good reasons. 

In this paper, however, we are not actually reducing any theory. We are simply organizing the ten-dimensional equations for	supersymmetry from a four-dimensional perspective. The gravitino multiplets will be for us a bookkeeping device; some of our equations will be in the ``vector representations'', and we now know that some will resemble those in a vector multiplet, while others will resemble the supersymmetry equations for a gravitino multiplet.

We finally want to understand whether the partners of the scalars (\ref{eq:sc31}) come from (\ref{eq:Emix}) or from (\ref{eq:F231}). To see this, it is again useful to think about the ${\cal N}=8$ theory. This theory has ${8 \choose 2}=28$ vectors, whose field-strengths we will denote by $T_{AB}$, antisymmetric in $AB$, and ${8 \choose 4}=70$ scalars parameterizing a coset space, whose vielbein we will denote by a totally antisymmetric $P_{ABCD}$.  In ${\cal N}=2$ terms, the index $A$ should be split in ${\rm SU}(3) \times {\rm SU}(3)$ representations. The first four $\zeta_A$ come from $\epsilon_1$, while the second four come from $\epsilon_2$. Taking also chirality into account, we see that
\begin{equation}
	\label{eq:Asplit}
	A \to (\mathbf{1}, \mathbf{1}) \oplus (\mathbf{3},\mathbf{1})  \oplus
	(\mathbf{1}, \mathbf{\bar 3}) \oplus (\mathbf{1}, \mathbf{1})\ .
\end{equation}
As in the generalized Hodge diamond (\ref{eq:hodge}), we can introduce an index $i_1$ for the $(\mathbf{3},\mathbf{1})$ and an index $\bar j_2$ for the $(\mathbf{1}, \mathbf{\bar 3})$. The first and second singlet will be denoted by indices $1$ and $2$. In this language, the RR vectors should be associated to field-strengths which mix indices coming from the first copy of SU(3) (namely, $1$ and $i_1$) with indices from the second copy ($2$ and $\bar j_2$):
\begin{equation}
	{\rm RR:}\quad T_{1 \bar j_2}\  (\mathbf{1}, \mathbf{\bar 3})\ ,
	T_{12}\ (\mathbf{1}, \mathbf{1}) \ , 
	T_{i_1 \bar j_2} \ (\mathbf{3}, \mathbf{\bar 3}) \ ,
	T_{i_1 2}\ (\mathbf{3}, \mathbf{1})\ .
\end{equation}
Clearly, $T_{12}$ is the graviphoton, $T_{i_1 \bar j_2}$ are the vectors in (\ref{eq:G+}), and $T_{1 \bar j_2}$, $T_{i_1 2}$ are the vectors from (\ref{eq:F231}). On the other hand, the following vectors should come from the NSNS sector:
\begin{equation}
	{\rm NSNS:} \quad T_{1 i_1}\  (\mathbf{3}, \mathbf{1})\ ,
	T_{i_1 j_1}\ (\mathbf{\bar 3}, \mathbf{1}) \ , 
	T_{\bar j_1 \bar j_2} \ (\mathbf{1}, \mathbf{3}) \ ,
	T_{\bar j_2 2}\ (\mathbf{1}, \mathbf{\bar 3})\ .
\end{equation}
Turning now to the scalars $P_{ABCD}$ of the ${\cal N}=8$ theory, in ${\cal N}=2$ terms we see that the $2\times {6 \choose 3}=40$ scalars which have one ``singlet'' index (1 or 2) sit in hypermultiplets, while the ${6 \choose 2}+ {6 \choose 4}=30$ which have both 1 and 2, or neither, sit in vector multiplets. The supersymmetry transformations of the spin $1/2$ fields $\lambda_{IJK}$ in ${\cal N}=8$ supergravity \cite[Sec.~7]{cremmer-julia} schematically read, in absence of gauging, $\delta \lambda_{ABC}\sim T_{[AB} \zeta_{C]+}+ P_{ABCD} \zeta^D_-$. In a ${\cal N}=2$ truncation, we only have the supersymmetry parameters $\zeta_1$ and $\zeta_2$, and we set to zero $\zeta_{i_1}$, $\zeta_{\bar j_2}$. Under ${\cal N}=2$ supersymmetry, then, $T_{i_1 j_1}$ is for example mixed with $P_{1 i_1 j_1 2}$, while $T_{1 \bar j_2}$ is not related to any $P$ (since $P_{1 1 \bar j_2 2}$ vanishes by antisymmetry). In other words, only the $T$ which do not have an index 1 or 2 can sit in a vector multiplet: 
\begin{equation}
	T_{i_1 \bar j_2}\ ,\qquad T_{i_1 j_1}\ ,\qquad T_{\bar j_1 \bar j_2} \qquad 
	{\rm (in\ vector\ multiplets)\ .}
\end{equation} 
This means that the vector which is a partner of (\ref{eq:sc31}) is among (\ref{eq:Emix}), rather than (\ref{eq:F231}). The multiplet structure is summarized in table \ref{t:multiplets}.  

\begin{table}[h]
\begin{center}
\begin{tabular}{|c|c|}\hline
	 scalars & vectors \\
\hline\hline
$\left.
\begin{array}{c}
\delta g_{mn}\\	\delta B_{mn}
\end{array}
\right\}
\sim \delta \phi_\pm \to \underbrace{(\mathbf{3}, \mathbf{3})}_{\rm hm} \oplus \underbrace{(\mathbf{3}, \mathbf{\bar 3})}_{\rm vm} $ & 
$\left.
\begin{array}{c}
\delta g_{\mu m}\\	\delta B_{\mu m}
\end{array}
\right\}
\to 
\underbrace{(\mathbf{3},\mathbf{1}) \oplus (\mathbf{1},\mathbf{\bar 3})}_{\rm gravitino\ mult.} \oplus
\underbrace{(\mathbf{\bar 3},\mathbf{1}) \oplus (\mathbf{1},\mathbf{3})}_{\rm vm}
$
\\ \hline
$F \to \underbrace{(\mathbf{1}, \mathbf{1})}_{\rm univ.\ hm} \oplus
\underbrace{(\mathbf{3}, \mathbf{3})}_{\rm hm} \oplus 
\underbrace{(\mathbf{\bar 3}, \mathbf{1}) \oplus (\mathbf{1}, \mathbf{3})}_{\rm vm} 
$
&
$
F \to \underbrace{(\mathbf{1}, \mathbf{1})}_{\rm graviphoton} \oplus
\underbrace{(\mathbf{3}, \mathbf{\bar 3})}_{\rm vm} \oplus 
\underbrace{(\mathbf{3}, \mathbf{1}) \oplus (\mathbf{1}, \mathbf{\bar 3})}_{\rm gravitino\ mult.}$ \\ \hline	
\end{tabular}
\end{center}
\caption{Summary of the multiplet structure.}
\label{t:multiplets}
\end{table}



\subsection{External gravitino equations} 
\label{sub:4dgeom}

We will now start collecting the ten-dimensional supersymmetry equations. We will start in this section by collecting those that constrain the four-dimensional geometry. As argued in section \ref{sub:timelike4d}, the information contained in the covariant derivatives $\nabla_\mu \zeta_i$ can be completely extracted from the exterior derivatives $d\mu$, $d k_i$, $dv$. 

The exterior derivatives $d \mu$, $d v^3$, $d v$ can be extracted from section \ref{sec:10d}, while $dk$ will have to be rederived. (Recall that $k=\frac12 (k_1 + k_2)$, $v^3 = \frac12 (k_1 - k_2)$.) The equation for $d v^3$ was given in (\ref{eq:dKtilde2}), and we repeat it here for convenience:
\begin{subequations}
\begin{equation}\label{eq:dtk10}
	d_4 v^3 = \iota_k H_3  \ .
\end{equation}
$d\mu$ and $dv$ can be obtained by taking the pairing of (\ref{eq:ext1}) with $\bar\phi_+$ and of (\ref{eq:ext2}) with $\phi_-$ respectively (recall also the relations (\ref{eq:normpurespinor})): 
\begin{align}
	\label{eq:dmu10}
	&d_4 \mu - i \mu\, (\bar \phi_+ , e^\phi d_4 (e^{-\phi} \phi_+)) = 
	s^x v^x - 2 \iota_k T^+  \ ,\\ 
	\label{eq:dv10}
	&  d_4 v +2i v \wedge  (\bar \phi_- , e^\phi d_4 (e^{-\phi} \phi_-)) = - \frac 12 \omega \bar{s}^- + \frac 12\bar{\omega} s^+ - i  e^\phi \iota_k (\bar{\phi}_-, F_3) - i e^\phi v^3 \wedge (\bar{\phi}_-, F_1) 
\end{align} 
\end{subequations}
where $T^+$ was defined in (\ref{eq:t+}), and
\begin{equation}\label{eq:si}
	s^+ =  4i\,(\bar \phi_+ , d_6 \bar \phi_-) \ ,\qquad
	s^- =  4i\,(\bar \phi_+ , d_6 \phi_-) \ ,\qquad
	s^3 =  i e^\phi\,(\bar \phi_+, F_0)\ .
\end{equation}
These $s_i$ are morally related to the Killing prepotentials ${\cal P}_x$ of ${\cal N}=2$ supergravity; this identification agrees with \cite[Eq.~(2.139),(2.140)]{grana-louis-waldram} (up to a change in conventions). Notice that (\ref{eq:dmu10}) takes exactly the same form of the four-dimensional counterpart (\ref{eq:timesusy}) by simply putting
\begin{equation}\label{twisteddiff10}
  D \mu \equiv d_4 \mu - i \mu\, (\bar \phi_+ , e^\phi d_4 (e^{-\phi} \phi_+)) \ ,
\end{equation}
this result suggests the identification
\begin{equation}\label{Qhat10}
  \hat{Q} = - (\bar \phi_+ , e^\phi d_4 (e^{-\phi} \phi_+)) \ .
\end{equation}

Finally, we have to compute an expression for $dk$. Such an equation is not explicitly present in the original system of equations (\ref{eq:susy10}). In appendix \ref{sub:necpairing} we show that this equation would originate from the pairing equations (\ref{eq:++1}), (\ref{eq:++2}); however, we found it easier to compute it from scratch. We start from the equation 
\begin{equation}\label{eq:dK10d}
  D_{[M} K_{N]} - \frac{1}{2}\, H_{MNQ} \tilde{K}^Q = - \frac{e^\phi}{256} \, \bar{\epsilon}_1 \Gamma_{[M|} F\, \Gamma_{|N]} \epsilon_2\ ,
\end{equation}
which is valid in ten dimensions without making any assumption about compactifications.  To specialize (\ref{eq:dK10d}) to compactifications, recall that $k$ and $v^3$ have only external components (and that they only depend on the external coordinates). Using the decompositions of spinors and fluxes in sections \ref{sub:factor} and \ref{sub:flux}, one finally obtains
\begin{equation}\label{eq:dkfinal}
  d_4 k - \iota_{v^3} H  = 2 \mathrm{Re} \left[-  \bar{\omega} \,s_3 + e^\phi(\bar{\phi}_-, F_1) \llcorner (\ast \bar{v}) - 2\bar{\mu}\, T^+ \right] \ .
\end{equation}

Together, (\ref{eq:dtk10}), (\ref{eq:dmu10}), (\ref{eq:dv10}) and (\ref{eq:dkfinal}) exhaust the constraints on the geometry of the external spacetime $M_4$. They are the analogues of (\ref{eq:timesusy}) for four-dimensional ${\cal N}=2$ supergravity.


\subsection{Universal hypermultiplet} 
\label{sub:univ}

As we saw in section \ref{sub:org}, the scalars in the universal hypermultiplet are the dilaton $\phi$, the axion $a$ defined in (\ref{eq:da}), and the complex scalar $\xi$ defined by (\ref{eq:Dxi}).

The equations for these scalars are not easy to find in the system we gave in section \ref{sec:10d}. They are hidden inside some equations that would seem to constrain four-dimensional geometry. There are several of these equations: the equation for $d_4 \omega$ in (\ref{eq:ext3}) and for $d_4 * v$ in (\ref{eq:ext4}); (\ref{eq:Kkilling2}), saying that $K_\mu$ is a Killing vector for $M_4$; and the `pairing' equations (\ref{eq:++1q}), (\ref{eq:++1p}), (\ref{eq:++2q}), (\ref{eq:++2p}). One can eliminate from these equations all the four-dimensional intrinsic torsions by using (\ref{eq:dtk10}), (\ref{eq:dmu10}), (\ref{eq:dv10}) and (\ref{eq:dkfinal}). Some of the equations become trivial; four stay non-trivial, and can be interpreted as the equations for the scalars in the universal hypermultiplet. 

This derivation is laborious, however, and in this case it might be preferable to present an alternative logic, stemming directly from the supersymmetry equations.	

The idea is to start from the dilatino equations in ten dimensions. These read $(-\frac12 H + \del \phi) \epsilon_1 + \frac{e^\phi}{16} \Gamma^M F \Gamma_M \epsilon_2=0$, $(\frac12 H + \del \phi) \epsilon_2 + \frac{e^\phi}{16} \Gamma^M \lambda(F) \Gamma_M \epsilon_1 = 0$. They do not contain any intrinsic torsions, either internal or external (unless one defines the intrinsic torsions by including the NSNS flux, as was done in \cite[App.~B]{10d}). One way to massage it is to use $\gamma^\mu C_k \gamma_\mu = (-)^k (4-2k) C_k$, where $C_k$ is $k$-form in four dimensions, and write
\begin{equation}
	\Gamma^M F \Gamma_M = 2 (4 F_0 + 2 F_1) + \Gamma^m F \Gamma_m \ . 
\end{equation}
The last term can be eliminated using the internal gravitino, which in turn produces a term involving the internal Dirac operator $\gamma^m D_m$. Taking the inner product of the spinorial equation thus obtained with the spinors in (\ref{eq:46basis}) (and its analogue with ${}_1 \to {}_2$) produces several equations. Some have an internal free index, and naturally belong to the ``edge of the diamond'' equations that we will present in sections \ref{sub:edge-vm} and \ref{sub:edge-gino}. The ones without an internal index are
\begin{subequations}\label{eq:univ}
	\begin{align}
		&L_K \phi = 0 \ ,\qquad L_K a = - 4 {\rm Im} (\mu \bar s^3)\ ,\qquad	L_K (\xi + i \tilde \xi) =  i ( \mu \bar{s}^- - \bar \mu s^+ )\ ;\\
		& v^3 \cdot d a + {\rm Re}( v \cdot D(\xi - i \tilde \xi) ) = 0 \ ,\quad
		-2 v^3 \cdot d\phi +  {\rm Im}( v \cdot D(\xi - i \tilde \xi) ) =4 {\rm Re} (\mu \bar s^3)\ , \\
		&v^3 \cdot D (\xi + i \tilde \xi) - v \cdot d(a + 2i \phi) =   i (\mu \bar s^- + \bar \mu s^+) \ .
	\end{align} 	
\end{subequations}
These can be written exactly as (\ref{eq:timesusyhyper}): it is enough to define
\begin{equation}\label{eq:uhyper}
	q^1 = a \ ,\qquad q^2 = \xi \ ,\qquad q^3 = \tilde\xi \ ,\qquad
	q^4 = 2 \phi \ ,
\end{equation}
to take the matrices $\Omega^x$ to be
\begin{equation}\label{eq:Ki10}
  \Omega^1 = \left(\begin{array}{cccc}
		0 & 0 & 1 & 0 \\
		0 & 0  & 0 & 1 \\
		-1 & 0 & 0  & 0 \\
		0 & -1 & 0 & 0 
	\end{array}\right) \ ,\qquad
	\Omega^2 = \left(\begin{array}{cccc}
		0 & -1 & 0 & 0 \\
		1 & 0  & 0 & 0 \\
		0 & 0 & 0 & 1 \\
		0 & 0 & -1 & 0 
	\end{array}\right) \ ,\qquad
	\Omega^3 = \left(\begin{array}{cccc}
		0 & 0 & 0 & 1 \\
		0 & 0  & -1 & 0 \\
		0 & 1 & 0 & 0 \\
		-1 & 0  & 0 & 0 
	\end{array}\right)\ .
\end{equation}
and the gauging vectors 
\begin{equation}\label{eq:gaugvu}
	g\,\bar{{\cal L}}^\Lambda k^u_\Lambda = 
	\left(\begin{array}{c}
	- 4\bar{s^3} \\
	-(\bar{s}^+ +  \bar{s}^-) \\
	i(\bar{s}^- - \bar{s}^+) \\
	0
	\end{array}	 \right) \ .
\end{equation}
The $\Omega^x$ in (\ref{eq:Ki10}) can be thought of as a block of dimension 4 of the triplet of complex structures $\Omega^x$ (which appeared in section \ref{sub:4dhyno}) characterizing the quaternionic structure of the space of fields of a four-dimensional ${\cal N}=2$ supergravity. Notice that indeed (\ref{eq:Ki10}) satisfy $\Omega^x \Omega^y = - \delta^{xy} 1 + \epsilon^{xyz} \Omega^z$. 


\subsection{Non-universal hypermultiplets} 
\label{sub:otherhyp}

We will now present the equations corresponding to the other, non-universal, hypermultiplets. As we saw in section \ref{sub:org}, the scalars are the $z^\alpha$, $\tilde z^\alpha$ defined in (\ref{eq:Dz}), which come from deformations of the internal metric along the forms (\ref{eq:dpal-}), and the $\xi^\alpha$, $\tilde \xi^\alpha$ defined in (\ref{eq:Dxi}), which come from fluxes. 

It is natural to look for these equations in (\ref{eq:ext0}), (\ref{eq:ext2}), (\ref{eq:ext4}), by taking their pairing with the forms $\delta \phi^\alpha_-$ in (\ref{eq:dpal-}). However, it turns out that the pairings $(\delta \phi^\alpha_-,(\ref{eq:ext0}))$ and $(\delta \phi^\alpha_-,(\ref{eq:ext4}))$ are redundant: they give equations that can also be obtained from $(\delta \phi^\alpha_-,(\ref{eq:ext2}))$. We hence need to consider only the latter pairing. Using the self-duality property (\ref{eq:F*F}), one can show that
\begin{equation}
	(\delta \phi^\alpha_-, F_3) = - i * (\delta \phi^\alpha_-, F_1) \ .
\end{equation}
This allows to rewrite $(\delta \phi^\alpha_-,(\ref{eq:ext2}))$ as
\begin{equation}\label{eq:2formhyper}
	v \wedge (\delta \phi^\alpha_-, d_4 \phi_-) - (v^3 - i \iota_k *) D(\xi + i \tilde \xi) = - \omega (\delta \phi^\alpha_-, d_6 \phi_+ ) - \bar \omega (\delta \phi^\alpha_-, d_6 \bar \phi_+) \ .
\end{equation}
This equation becomes more familiar once we decompose it along the basis of two-forms
$\{ k \wedge v^x, v^x \wedge v^y\}$. The components along $k \wedge v^x$ give the equations
\begin{subequations}\label{eq:nonuniv}
\begin{align}
	\label{eq:Lkza}
	&L_k z^\alpha = L_k \tilde z^\alpha = 0\ , \\ 
	\label{eq:Lkxia}
	&L_k D( \xi + i \tilde \xi) = \mu (\delta \phi^\alpha_-, d_6 \phi_+) + \bar \mu (\delta \phi^\alpha_-, d_6 \bar \phi_+) \ .
\end{align}
(\ref{eq:Lkxia}) comes from $k\wedge v^3$, and is exactly the same as $(\delta \phi^\alpha_-,(\ref{eq:ext0}))$; (\ref{eq:Lkza}) come from $k\wedge v^1$, or equivalently $k \wedge v^2$. Notice that (\ref{eq:Lkza}) also follow from the statement (\ref{eq:Kkilling0}), that the action of $k$ preserves the internal metric $g_{mn}$. The components of (\ref{eq:2formhyper}) along $v^x \wedge v^y$ give
\begin{align}
		\label{eq:vDxia}
		&v \cdot D (\xi^\alpha + i \tilde \xi^\alpha) + v^3 \cdot D (z^\alpha + i \tilde z^\alpha) = 0 	\ ,\\
		\label{eq:vDza} 
		&v \cdot D (z^\alpha + i \tilde z^\alpha) - v^3 \cdot D (\xi^\alpha + i \tilde \xi^\alpha) = - \mu ( \delta \phi^\alpha_-, d_6 \phi_+) + \bar \mu ( \delta \phi^\alpha_-, d_6 \bar \phi_+)\ .
\end{align}
(\ref{eq:vDza}) comes from $v^1 \wedge v^2$, and is exactly the same as $(\delta \phi^\alpha_-,(\ref{eq:ext4}))$. On the other hand, (\ref{eq:vDxia}) comes from $v^1 \wedge v^3$, or equivalently from $v^2 \wedge v^3$. In black hole applications, these equations would often give that hypermultiplet scalars do not flow, but sometimes that they do, as in \cite{halmagyi-petrini-zaffaroni-bh}. 
\end{subequations}

All the equations in (\ref{eq:nonuniv}) can again be rewritten as in (\ref{eq:timesusyhyper}) by taking
\begin{equation}\label{eq:nuhyper}
 q^{1 \alpha} = z^\alpha \qquad q^{2 \alpha} = \xi^\alpha \qquad q^{3 \alpha} = \tilde{\xi}^\alpha \qquad q^4 = \tilde{z}^\alpha
\end{equation} 
and the $\Omega^{x}$, in each dimension 4 block corresponding to a hypermultiplet, are again given by (\ref{eq:Ki10}).
The gauging vectors are given by 
\begin{equation}\label{eq:gaugvnu}
	g \bar {\cal L}^\Lambda k^u_\Lambda = \left( \begin{array}{c}
		0  \\
		i (\delta \phi^\alpha_-, d_6 \phi_+)+i (\overline{\delta \phi^\alpha_-}, d_6 \phi_+) \\
		(\delta \phi^\alpha_-, d_6 \phi_+) - (\overline{\delta \phi^\alpha_-}, d_6 \phi_+) \\
		0
	\end{array}\right)\ .
\end{equation}


\subsection{Vector multiplets} 
\label{sub:vm}

Now we turn to the equations corresponding to the vector multiplets. As we saw in section \ref{sub:org}, the scalars in the vector multiplets sit in the components along $\delta \phi^a_+$ in (\ref{eq:dpal+}) and its conjugate. The pairing 
$(\delta \phi^a_+, (\ref{eq:ext1}))$ gives
\begin{equation}\label{eq:vector10d}
  - i \mu\, D \bar{t}^a +\frac i2 \bar{W}^{ax}v^x = 2 \iota_k G^{a -} \ ,
\end{equation}
where the $D \bar{t}^a$ and $G^{a-}$ were defined in (\ref{eq:Dt}), (\ref{eq:G+}), and the $W$'s are defined as
\begin{equation}\label{eq:defvectorscalars}
   \bar{W}^{a+} \equiv  4\,(\delta \phi_+^a , d_6  \phi_-) \, , \quad \bar{W}^{a-} \equiv  4\,(\delta \phi_+^a , d_6 \bar{\phi}_-) \, , \quad \bar{W}^{a3} \equiv e^\phi\,(\delta \phi_+^a, F_0)\ .  
\end{equation}
(\ref{eq:vector10d}) is exactly the complex conjugate of (\ref{eq:timesusygauginooneform}). The $W^{a\,x}$ correspond to the four-dimensional shifts appearing in (\ref{eq:4dgaugino}) and (\ref{eq:timesusygauginooneform}). They are related to the derivatives of the shifts $s^x$ with respects to the geometrical moduli. This feature was already remarked in \cite[Eq.~(3.77)]{cassani-bilal}. 

It is also interesting to note that there does not seem to exist an equivalent of the shift $W^a$. In fact, thanks to this, we see that 
\begin{equation}\label{eq:Lkta}
	L_k t^a=0\ .
\end{equation}
Another notable consequence of (\ref{eq:vector10d}) would be, in black hole applications, the attractor equation for vector multiplet scalars. 

It remains to consider the parings of (\ref{eq:ext3}) with $\delta \phi_+^a$. However one can show that the resultant equations are exactly equivalent to (\ref{eq:vector10d}), therefore they do not give any additional information. This result is not a surprise since we already know that (\ref{eq:timesusygaugino}) is sufficient to fully reconstruct the gaugino equations in the four-dimensional timelike case.


\subsection{New vector multiplets: edge of the diamond} 
\label{sub:edge-vm}

As we saw in section \ref{ssub:multiplets} (see the summary in table \ref{t:multiplets}), we also expect equations associated to the ``edge of the diamond''. They come from different places: to start with we have the equations coming from (\ref{eq:ext}) and expanded along the edge of diamond. These are similar to the counterparts just discussed in the interior of the diamond; however, in this case we have fewer redundacies than in sections \ref{sub:otherhyp} and \ref{sub:vm}, and thus the full amount of equations is bigger. We have also the pairing equations (\ref{eq:pairingf1}), (\ref{eq:pairingf2}), (\ref{eq:pairingf3}) and (\ref{eq:pairingf4}).\footnote{Recall that the other pairing equations (\ref{eq:++1q}), (\ref{eq:++1p}), (\ref{eq:++2q}) and (\ref{eq:++2p}) are redundant in the timelike case since they determines external torsions which are determined by the equations in section \ref{sub:4dgeom}.} All these equations will be shown in this section and in the next. 

In this section, we will give the ones corresponding to the new vector multiplet in the $(\mathbf{\bar 3}, \mathbf{1}) \oplus (\mathbf{1}, \mathbf{3})$ (see table \ref{t:multiplets}).  They take the form 
\begin{subequations}\label{eq:edge-vm}
\begin{align}
	\mu (\gamma^{\bar{i}_1} \phi_+, F_1) = 2 e^{-\phi} (\gamma^{\bar{i}_1} \phi_+, d_6 \bar{\phi}_+) (k - v^3) + (\gamma^{\bar{i}_1} \phi_-, F_0) v - 2 \left(\frac{e^{-\phi}}{\bar{\mu}} v \cdot (\gamma^{\bar{i}_1} \phi_-, d_4 \phi_+) \right) v^3\ ,\\
	\bar\mu (\bar\phi_+ \gamma^{\bar i_2}, F_1) = 2 e^{-\phi} (\bar\phi_+ \gamma^{\bar i_2}, d_6 \phi_+) (k + v^3) + (\phi_- \gamma^{\bar i_2}, F_0) v + 2 \left(\frac{e^{-\phi}}{\mu} v \cdot (\bar\phi_+ \gamma^{\bar i_2}, d_4 \phi_-) \right) v^3\ .
\end{align} 
\end{subequations}
They come from (\ref{eq:ext2}) expanded along the edge of the diamond, combined with (\ref{eq:pairingf2}) and (\ref{eq:pairingf4}). (Equations (\ref{eq:ext0}) and (\ref{eq:ext4}) are redundant like it happened for the non universal hypermultiplet  in section \ref{sub:otherhyp}).
The reason we do not see any field-strengths in the vector multiplet equations (\ref{eq:edge-vm}) is that their gauge potentials would be $g_{\mu m}$ and $B_{\mu m}$ (see again table \ref{t:multiplets}), which we have set to zero. 


\subsection{Gravitino multiplet: edge of the diamond} 
\label{sub:edge-gino}

We finally show the equations associated to the gravitino multiplet of table \ref{t:multiplets}. This was not discussed in section \ref{sec:4dsugratime}, since it is usually not considered in four-dimensional theories, for reasons explained in section \ref{ssub:multiplets}. As discussed there, we are not advocating including gravitino multiplets in a reduction; in this paper, we are not reducing, but rather organizing the ten-dimensional supersymmetry equations in a way which makes it natural to compare with four dimensions. 

The equations read
\begin{subequations}\label{eq:ginoeq}:
	\begin{align}
	&\iota_k (\gamma^{i_1} \bar\phi_-, F_2) = - 2 \bar\mu e^{-\phi} (\gamma^{i_1} \bar\phi_-, d_4 \bar\phi_+) - 2 Q^x_L v_x  \label{eq:ginoeq1}	\\
	 &\iota_k (\phi_- \gamma^{\bar i_2}, F_2) = - 2 \bar\mu e^{-\phi} (\phi_- \gamma^{\bar i_2}, d_4 \bar\phi_+) - 2 Q^x_R v_x  \label{eq:ginoeq2}\\
	 &\iota_{v^3} (\gamma^{\bar{i}_1}\phi_-, d_4 \phi_+) = \iota_{v^3} (\phi_- \gamma^{\bar i_2}, d_4 \bar\phi_+) = 0 = \iota_{k} (\gamma^{\bar{i}_1}\phi_-, d_4 \phi_+) = \iota_{k} (\phi_- \gamma^{\bar i_2}, d_4 \bar\phi_+)  \ , \label{eq:ginoeq3}\\
	 & (\gamma^{\bar{i}_1} \phi_+, d_6 \bar{\phi}_+) = (\gamma^{\bar{i}_1} \phi_-, d_6 \bar{\phi}_-)
	\ ,\qquad
	 (\bar\phi_+ \gamma^{\bar i_2}, d_6 \phi_+) = - (\phi_- \gamma^{\bar i_2}, d_6 \bar{\phi}_-) \ ,\label{eq:ginoeq4}
	\end{align}	
\end{subequations}
where 
\begin{equation}\label{Q^x}
	Q^x_L = \left(\begin{array}{c}
		e^{-\phi} (\gamma^{i_1} \bar{\phi_-}, d_6 \phi_-) \\
		i e^{-\phi} (\gamma^{i_1} \bar\phi_-, d_6  \phi_-) \\
		 \frac 12 (\gamma^{i_1} \bar{\phi_-}, F_0) 
	\end{array}\right)
	\ ,\qquad
	Q^x_R = \left(
	\begin{array}{c}
		e^{-\phi} (\phi_- \gamma^{\bar i_2}, d_6 \bar \phi_-) \\
		-i e^{-\phi} (\phi_- \gamma^{\bar i_2}, d_6 \bar \phi_-) \\
		 \frac 12 (\phi_-\gamma^{\bar i_2}, F_0)
	\end{array}
	\right)\ .
\end{equation}
(\ref{eq:ginoeq1}) and (\ref{eq:ginoeq2}) are obtained by taking  $(\gamma^{\bar{i}_1} \phi_-, (\ref{eq:ext1}))$ and $(\phi_- \gamma^{\bar{i}_2}, (\ref{eq:ext1}))$, while (\ref{eq:ginoeq3}) are obtained from $(\gamma^{\bar{i}_1} \phi_-, (\ref{eq:ext2}))$ and $(\phi_- \gamma^{\bar{i}_2}, (\ref{eq:ext2}))$. (\ref{eq:ginoeq4}) are a consequence of (\ref{eq:pairingf1}) and of (\ref{eq:pairingf3}). 

From (\ref{eq:ginoeq3}), together with (\ref{eq:Lkza}) and (\ref{eq:Lkta}), we see that the internal pure spinors $\phi_\pm$ are left invariant by the action of $k$. Since, as we recalled at the beginning of section \ref{ssub:scalars}, the internal metric and $B_0$ field are determined by them, we recover (\ref{eq:Kkilling0}), (\ref{eq:dKtilde0}).

With all the caveats given above, we can use our rough discussion in section \ref{ssub:multiplets} to get a sense of where the new equations (\ref{eq:ginoeq}) come from. In ${\cal N}=8$ supergravity, in absence of gaugings the supersymmetry transformations of gravitinos look like $\delta \psi_{\mu\,A+}\sim D_\mu \zeta_{A+} + T^+_{AB} \gamma_\mu \zeta^B_-$. Recall that for us the index $A$ is split in $1, i_1, \bar j_2, 2$ (see (\ref{eq:Asplit})). Moreover, only $\zeta_1$ and $\zeta_2$ are kept, while the remaining $\zeta_{i_1}$ and $\zeta_{\bar j_2}$ are set to zero. The gravitino equations for $A=1,2$ now become the usual ${\cal N}=2$ gravitino equations (\ref{eq:4dgrino}) (corrected there by gaugings); for $A=i_1,\bar j_2$ they become the new equations
\begin{equation}
	\label{eq:Teqgr}
	T_{i_1 1 +}^{\mu \nu} \gamma_\nu \zeta^1 + T_{i_1 2 +}^{\mu \nu} \gamma_\nu \zeta^2 = 0  =
	T_{\bar j_2 1 +}^{\mu \nu} \gamma_\nu \zeta^1 + T_{\bar j_2 2 +}^{\mu \nu} \gamma_\nu \zeta^2\ .
\end{equation}
In ${\cal N}=8$ we also have the supersymmetry variations of the spin 1/2 fields, which again read $\delta \lambda_{ABC}\sim T_{[AB} \zeta_{C]+}+ P_{ABCD} \zeta^D_-$.  Now, $\delta\lambda_{i_1 \bar j_2 1}$ and $\delta\lambda_{i_1 \bar j_2 2}$ give rise to the vector multiplet equations in \ref{sub:vm}; $\delta\lambda_{i_1 j_1 1}$ and $\delta\lambda_{i_1 j_1 2}$ give rise to the ``edge'' vector multiplets discussed in section \ref{sub:edge-vm}; the $\delta \lambda$'s with no 1 or 2 index give rise to the hypermultiplets, discussed in section \ref{sub:univ} and \ref{sub:otherhyp}. We still have $\delta \lambda_{i_1 12}$ and $\delta \lambda_{\bar j_2 12}$; these give 
\begin{equation}
	\label{eq:Teqdil}
	T_{i_1 [1} \zeta_{2]} = 0 = T_{\bar j_2 [1} \zeta_{2]}\ .
\end{equation}
Both (\ref{eq:Teqgr}) and (\ref{eq:Teqdil}), suitably corrected by gaugings, are the origin of (\ref{eq:ginoeq}).


\subsection{Nontrivial fibrations} 
\label{sub:fibr}

The metric Ansatz (\ref{eq:metAn}) we have followed so far did not have any mixed metric components $g_{\mu m}$; in other words, we have assumed that the fibration is topologically trivial. As remarked in the introduction (see footnotes \ref{foot:fibr} and \ref{foot:bh}), this is enough for many applications, but not for example for some black holes. We will discuss here how one should proceed to extend our results to nontrivial fibrations. Although we leave detailed computations to further research, we will argue that under natural assumptions its effects are limited to the ``edge'' multiplets of sections \ref{sub:edge-vm} and \ref{sub:edge-gino}.

In a topologically nontrivial fibration, the metric would read 
\begin{equation}\label{eq:nontrfibr}
	ds^2_{10} = ds^2_4(x) + g^6_{mn} (dy^m + K^m) (dy^n + K^n) \ ,\qquad
	K^m = K^m_\mu dx^\mu\ .
\end{equation}
The spacetime one-forms $K^m$ are interpreted as connections; if $M_6$ has nontrivial isometries, one usually takes the $K^m$ to be linear combinations of those isometries. The second term in (\ref{eq:nontrfibr}) replaces $ds^2_6=g^6_{mn} dy^m dy^n$ in (\ref{eq:ds46}). In our previous analysis, then, in all our internal forms we should replace
\begin{equation}
	dy^m \to dy^m + K^m \ .
\end{equation}
This can be achieved by acting with the operator 
\begin{equation}
	\iota_! \equiv e^{K\circ} \ ,\qquad K\circ \equiv K^m_\mu dx^\mu \iota_m \ . 
\end{equation}
For example, if $\alpha_k= \frac1{k!}\alpha_{m_1 \ldots m_k} dy^{m_1}\wedge \ldots \wedge dy^{m_k}$ is an internal $k$-form, we have $\iota_! \alpha_k = \frac1{k!}\alpha_{m_1 \ldots m_k} (dy^{m_1}+K^{m_1})\wedge \ldots \wedge (dy^{m_k} + K^{m_k})$. 

One should now act with this operator on our previous computations; in particular, $\Phi$ in (\ref{eq:Phi46}) should be now replaced by $\iota_! \Phi$. This will now change the rest of our computations, since now the action of $d = d_4 + d_6 = dx^\mu \del_\mu + dy^m \wedge \del_m$ is complicated by the presence of the $K^m_\mu$. 

Fortunately, it is possible to repackage all the new terms rather compactly. Since $d( \iota_! \alpha_k) = d (e^{K\circ} \alpha_k) = e^K (e^{-K\circ} d e^{K\circ}) \alpha_k = \iota_!(e^{-K\circ} d e^{K\circ}) \alpha_k$, it is enough to compute the action of the operator $e^{-K\circ} d e^{K\circ}$. This can be done using the ``Hadamard lemma'' $e^{-B} A e^B= A + [A,B]+ \frac12 [[A,B],B] +\ldots$. The first term is 
\begin{equation}
	[d,K\circ] = (d_4 K^m) \iota_m - dx^\mu \wedge \{ d_6, K^m_\mu \iota_m \} = 
	(d_4 K^m) \iota_m - dx^\mu \wedge L_{K_\mu} \ ,\qquad K_\mu \equiv K^m_\mu \del_m\ .
\end{equation}
The second term is 
\begin{equation}
	[[d,K\circ],K\circ] = -dx^\mu \wedge [L_{K_\mu}, K\circ] = -dx^\mu \wedge dx^\nu \wedge [L_{K_\mu}, \iota_{K_\nu}]=  -dx^\mu \wedge dx^\nu \wedge \iota_{[K_\mu,K_\nu]}\ ,
\end{equation}
where $[K_\mu,K_\nu]$ is now simply the Lie bracket of the two internal vectors $K_\mu$ and $K_\nu$. Now $[[[d,K\circ],K\circ],K\circ]=0$, and the Hadamard series stops. We can then reassemble the terms to get 
\begin{equation}\label{eq:diota!}
	d \iota_! 
	= \iota_! (d - dx^\mu \wedge L_{K_\mu} +\iota_m F^{K\,m}\wedge)\ ,
\end{equation}
where
\begin{equation}
	 F^K \equiv \frac12 \left(\del_\mu K^m_\nu -\frac12[K_\mu,K_\nu]^m\right) dx^\mu \wedge dx^\nu\ .
\end{equation}

In a typical situation, as we said we would take the $K^m$ to be linear combinations of internal isometries. Most of these also annihilate the internal pure spinors $\phi_\pm$; only the isometries corresponding to internal R-symmetries do not. Let us assume for simplicity that we only need to include $K_\mu$ such that $L_{K_\mu} \phi_\pm=0$, so that the first term in (\ref{eq:diota!}) disappears.  

The $F^K$ term in (\ref{eq:diota!}) is satisfying because it contains the field strength for the vectors that we called $\delta g_{\mu m}$ in table \ref{t:multiplets}. If again the $K_\mu$'s are taken to be linear combinations of the internal isometries, of course the Lie bracket $[K_\mu,K_\nu]$ would close on the isometries themselves. The effect of this term on the equations we have derived in this section is relatively mild. Since this term contains a single $\iota_m$, it will disappear from the pairings with most of the forms in the diamond (\ref{eq:hodge}), namely from those with $\phi_\pm$ and with the ``bulk of the diamond'' $\delta \phi^a_+$ in (\ref{eq:dpal+}) and $\delta \phi^\alpha_-$ in (\ref{eq:dpal-}). This means that the equations in sections \ref{sub:4dgeom}, \ref{sub:univ}, \ref{sub:otherhyp}, \ref{sub:vm} will not change. 

On the other hand, the equations in sections \ref{sub:edge-vm} and \ref{sub:edge-gino} will be changed, and need to be recomputed. Notice that in table \ref{t:multiplets} we had indeed predicted that the effect of the field strengths $F^K$ should be felt in the ``edge'' vector multiplets and in the gravitino multiplets of those two sections. Notice that according to table \ref{t:multiplets} we also expect $H_2$ (namely, the component of $H$ with two spacetime indices) to contribute to these equations; recall that in this paper we have set it to zero (see (\ref{eq:H1H20})).

Rather than attempting to derive general equations here, however, we leave this task to later research.


\subsection{Summary: ten-dimensional timelike case} 
\label{sub:10dtimelikesummary}

In this section, we refined our results of section \ref{sec:10d} for the timelike case. Namely, we still work with a fibration (\ref{eq:metAn}), with spinor Ansatz (\ref{eq:spAn}), (\ref{eq:eqnorms}) and assumptions summarized in footnote \ref{foot:fibr} and (\ref{eq:H1H20}), but now we assume that the spinors $\zeta_i$ do not coincide. We have found a system that is equivalent to preserved supersymmetry in this case. 
It consists of equations (\ref{eq:dtk10}), (\ref{eq:dmu10}), (\ref{eq:dv10}), (\ref{eq:dkfinal}); (\ref{eq:univ}), (\ref{eq:nonuniv}), (\ref{eq:vector10d}), (\ref{eq:edge-vm}), (\ref{eq:ginoeq}).\footnote{Let us compare more explicitly with section \ref{sub:10d}. Apart from equations (\ref{eq:dkfinal}) and (\ref{eq:univ}), which are obtained for the first time in this section, all the equations for the timelike case come from (\ref{eq:dKtilde2}), (\ref{eq:ext1}), (\ref{eq:ext2}), (\ref{eq:pairingf1}), (\ref{eq:pairingf2}), (\ref{eq:pairingf3}) and (\ref{eq:pairingf4}).} Most of the equations correspond to the ``boxed'' system in section \ref{sec:4dsugratime}; the ones which do not, (\ref{eq:ginoeq}), capture extra equations which it would be challenging to obtain in a four-dimensional effective approach, as anticipated in the introduction and discussed in detail in section \ref{ssub:multiplets}. We will discuss again the system in section \ref{sec:conc}.



\section{An aside: ${\cal N}=2$ four-dimensional supergravity, general case} 
\label{sec:4dgen}

We found in section \ref{sec:4dsugratime} a system of form equations which is equivalent to preserved supersymmetry in the timelike case. This was based on the fact (see section \ref{sub:timelike4d}) that all the intrinsic torsions can be reconstructed from $d\mu$, $dv$, $dk_i$.  This  had no clear counterpart in the general case, and at that point it was not clear how to proceed. 

In this section, we will see that the general ten-dimensional system we presented in section \ref{sec:10d} suggests a system of equations which are valid in the general case. Unfortunately, this system will be much more complicated than the one in section \ref{sec:4dsugratime}. It is possible that a better alternative exists; finding one might also suggest how to improve the ten-dimensional system in \cite{10d}. 

We will start in section \ref{sub:torsgen} by presenting a set of derivatives of forms from which the covariant derivatives of spinors can be fully reconstructed, in the general case. We will then describe our equations for the gravitinos (section \ref{sub:4dgravgen}), the hyperinos (section \ref{sub:4dhynogen}), and the gauginos (section \ref{sub:4dhgauginogen}).

\subsection{Intrinsic torsions in the general case} 
\label{sub:torsgen}

We begin by some mathematical preliminaries about how to reexpress the covariant derivatives $\nabla \zeta_i$ in terms of exterior algebra of forms. In the timelike case, this was done in \ref{sub:timelike4d}. In the general case, we have to use a different procedure; we will take inspiration from the intrinsic torsion for an $\rr^2$ structure defined by a single spinor \cite{ckmtz}, which we now review.

\subsubsection{One spinor} 
\label{ssub:inttors1}

A possible definition of intrinsic torsions for one Weyl spinor $\zeta$ was introduced in \cite{ckmtz}; we will briefly review it here.

As usual, the idea is to expand the derivative of the object defining the structure in terms of a basis; for example (\ref{eq:+basis}).  One can show using (\ref{eq:4dpure}) that
\begin{equation}\label{eq:ge+}
\gamma^\mu \zeta_+ = - w^\mu \zeta_- +\frac 1 2 k^\mu e^-\cdot \zeta_+ \ .
\end{equation}
We can also expand
\begin{equation}\label{eq:pq}
	\nabla_\mu \zeta_+ = p_\mu \zeta_+ + q_\mu e^- \cdot \zeta_-\ .
\end{equation}
$p_\mu$, $q_\mu$ are complex one-forms. 

The $\rr^2$ structure defined by $\zeta$ can also be described by
the one-forms $k$ and $w$, and indeed their exterior derivatives contain intrinsic torsions $p_\mu$ and $q_\mu$:
\begin{subequations}\label{eq:viel}
	\begin{align}
		d k &=  2 {\rm Re}\, p \wedge k + 4 {\rm Re}(q \wedge \bar  w)\ , \label{eq:dz}\\
		d w &=  - 2 \rho \wedge z + 2 i {\rm Im} p \wedge  w -2 q \wedge e^- \ .  \label{eq:dw}
	\end{align}	
$p= p_\mu dx^\mu $ and $q= q_\mu dx^\mu$ are the one-forms appearing in  (\ref{eq:pq}), while $\rho$ is a new one-form which is not an intrinsic torsion for the $\rr^2$ structure. (\ref{eq:dz}) and (\ref{eq:dw}) cannot be used to reconstruct the whole $p_\mu$ and $q_\mu$; as it turns out \cite{ckmtz}, one also has to add the information from $d e^-$:
\begin{equation}
	d e^-  =  4 {\rm Re}(\rho  \wedge \bar w) -2 {\rm Re} p \wedge e^- \ . \label{eq:de-} 
\end{equation}
\end{subequations}
Together, (\ref{eq:viel}) can now be used to reconstruct $p$ and $q$ completely. The one-forms $\rho$ appears because it is an intrinsic torsion for the identity structure $\{k, e^-, w, \bar w\}$ (together with $p$, $q$), even though it is not part of the intrinsic torsion for the $\rr^2$ structure defined by $\zeta_+$ (while $p$ and $q$ are). We also note for later use an expression for the codifferential of $e^-$:
\begin{equation}\label{eq:dde-}
	d^\dagger e^- = -2 e^-\cdot {\rm Re} \, p + 4  {\rm Re} (\bar w \cdot \rho)\ .
\end{equation}

Instead of considering $d e^-$, one might think it should be useful to consider the full covariant derivative $\nabla_\mu k_\nu$ rather than just its antisymmetrized version $dk$. After all, the symmetrized covariant derivative $\nabla_{(\mu} k_{\nu) }$ appears in many contexts in supergravity, especially in Lorentzian signature. One easily evaluates
\begin{equation}\label{eq:nablaztors}
	\nabla_\mu k_\nu = 2 {\rm Re} (p_\mu k_\nu + 2 \bar q_\mu w_\nu)\ .
\end{equation}
We see that this determines $q$ and ${\rm Re} p$, but that it does not say anything about ${\rm Im} p$. The components $k\cdot {\rm Im} p$ and $w \cdot {\rm Im} p$ can be determined from $dw$ (or $d \omega$, for that matter); however, $e^- \cdot {\rm Im} p$ will still have to be determined by a suitable projection of $d e^-$. 


\subsubsection{Two spinors} 
\label{ssub:inttors2}

We will now consider the case where we have two spinors $\zeta_i$ of positive chirality. 

As in section \ref{ssub:inttors1}, to define intrinsic torsions we need a basis of spinors. Recall from section \ref{sub:4d2eps} that the two $\zeta_i$ are independent in the timelike case (and so they can used to define a basis in this case), but that they are proportional in the null case. Between these two extremes one has a ``general'' case in which the two spinors $\zeta_i$ are parallel in some points and independent in others. The timelike case has already been dealt with in section \ref{sub:timelike4d}, with the result that one needs to compute the exterior derivatives of $\mu$, $k$, $v^x$. 

In the general case, however, we have to proceed differently; we will follow the procedure of section \ref{ssub:inttors1}. The $e^-_i$ introduced in section \ref{sub:4d2eps} now give us two possible bases for spinors, along the lines of (\ref{eq:+basis}), (\ref{eq:-basis}). These bases are related, of course; for example we compute\footnote{Notice that the coefficient $\frac{1}{2} e_1^- \cdot \bar{v}$ is precisely the same coefficient $\alpha(x)$ that we encountered in (\ref{eq:spinornull}).}
\begin{equation}\label{eq:zeta21}
	\zeta_2 = \frac12 e^-_1 \cdot \bar v\, \zeta_1 - \frac 12 \mu \,e^-_1 \cdot \zeta^1 \ 
\end{equation}
and analogously
\begin{equation}
  \zeta_1 = \frac12 e_2^-\cdot v\, \zeta_2 + \frac12 \mu \,e_2^- \cdot \zeta^2\ .
\end{equation}
However, we will most often keep the two bases separate. We can now simply introduce two sets of $p_\mu$ and $q_\mu$: 
\begin{equation}\label{eq:pqi}
	\nabla_\mu \zeta_i = p^i_\mu\, \zeta_i + q^i_\mu\, e^-_i \cdot \zeta^i\ 
\end{equation}
where the indices are not summed.

We should now try to identify a system of PDE in terms of exterior algebra, from which one can reconstruct both the $p^i$ and $q^i$ in (\ref{eq:pqi}), similarly to how (\ref{eq:dz}), (\ref{eq:dw}), (\ref{eq:de-}) reconstruct $p$ and $q$ in section \ref{ssub:inttors1}. 

Let us first see in this new language what happens in the timelike case. Here, the $e^-_i$ can be taken to be proportional to the $k_i$ themselves, as in (\ref{eq:e-time}). Also, (\ref{eq:zeta21}) simply becomes $\zeta_2 = -\frac12 \mu e^-_1 \zeta^1$; taking covariant derivatives of both sides, and reexpressing them in terms of the $p^i$ and $q^i$ as defined in (\ref{eq:pqi}), we get
\begin{equation}
	p_2 = - p_1 + \frac{d \mu}{\mu} \ ,\qquad q_2 = 2 \mu^2 \bar \rho_1\ .
\end{equation}
This tells us that $p_2$ and $q_2$ (and hence $\nabla \zeta_2$) are determined if we manage to determine $p_1$, $q_1$ and $\rho_1$. For this, however, we can follow the strategy reviewed in section \ref{ssub:inttors1} for a geometry defined by one spinor. As noted there, one can determine $p$, $q$ and $\rho$ from the equations for $dz$, $dw$ and $de^-$ in (\ref{eq:viel}). Applying this to our case means that $p_1$, $q_1$ and $\rho_1$ will be determined by the equations for $dk_1$, $d w_1$ and $d e^-_1$. However, we can now remember our identifications (\ref{eq:e-time}), (\ref{eq:vtime}): $w_1$ and $e^-_1$ are proportional to $v$ and $k_2$ respectively. Since the proportionality factors contain $\mu$, overall we need equations for the $d\mu$, $dk_i$, $dv$. This agrees with (\ref{eq:listtime}), once we recall the definition of the $v^x$ from (\ref{eq:vx}).

The general case is much more involved than the timelike case. We saw in section \ref{ssub:inttors1} that the system (\ref{eq:viel}) is necessary and sufficient to fully reconstruct intrinsic torsions for the case with one spinor. With two spinors, the most obvious procedure would be to ``double'' that computation, and consider the equations for $d k_i$, $d w_i$, $d e^-_i$. 

This is obviously sufficient to determine $p^i$ and $q^i$. However, it is not suited to the applications to supergravity we will pursue later in this section. The reason is that $d w_i$, $d e^-_i$ would again contain two `spurious' one-forms $\rho^i$ (just as in (\ref{eq:dw}), (\ref{eq:de-}) for one spinor), which would not be determined by supersymmetry, since they do not appear in the covariant derivatives $\nabla_\mu \zeta_i$. One might try to improve the system (\ref{eq:viel}) so that $\rho$ never appears (for example, by taking appropriate projections of $dw$ and $de^-$). This would certainly deserve further investigation. Here we will however follow a different approach.

First of all, (\ref{eq:psp10}) suggests that we consider the exterior derivative of $\zeta_1 \overline{\zeta_2}$  and $\zeta_1 \overline{\zeta_2^c}$. These are easy to compute: 
\begin{equation}\label{eq:de1e2}
	\begin{split}
		d(\zeta_1 \overline{\zeta^2}) &= (p_1 + \bar p_2)\wedge \zeta_1 \overline{\zeta^2} + q_1 \wedge e^-_1 \zeta^1 \overline{\zeta^2} - \bar q_2 \wedge \zeta_1 \overline{\zeta_2} e^-_2\ ,\\
		d(\zeta_1 \overline{\zeta_2}) &= (p_1 + p_2)\wedge \zeta_1 \overline{\zeta_2} + q_1 \wedge e^-_1 \zeta^1 \overline{\zeta_2} -  q_2 \wedge \zeta_1 \overline{\zeta^2} e^-_2\ .	
	\end{split}
\end{equation}
When $\zeta_1=\zeta_2$, these reduce respectively to (\ref{eq:dz}) and to the formula $d\omega= 2p\wedge \omega -2 q \wedge (z \wedge e^- + w \wedge \bar w)$, that one can obtain from (\ref{eq:dz}) and (\ref{eq:dw}). 

(\ref{eq:de1e2}) are not enough to determine $p_i$ and $q_i$. To see this, and to determine which other equations should supplement it, we should expand (\ref{eq:de1e2}) in an appropriate basis of forms. The most natural basis is given by considering forms as bispinors; a basis for bispinors can then be obtained by tensoring two copies of our spinorial bases. For example, a basis for even forms is given by tensoring two copies of (\ref{eq:+basis}), and two copies of (\ref{eq:-basis}):
\begin{equation}\label{eq:evenforms}
	\begin{split}
		\zeta_1 \overline{\zeta^2} \ ,\qquad
		e^-_1 \zeta^1 \overline{\zeta^2} \ ,\qquad
		\zeta_1 \overline{\zeta_2} e^2_- \ ,\qquad
		e^-_1 \zeta^1 \overline{\zeta_2} e^2_- \ ;\\
		\zeta^1 \overline{\zeta_2} \ ,\qquad
		e^-_1 \zeta_1 \overline{\zeta_2} \ ,\qquad
		\zeta^1 \overline{\zeta^2} e^2_- \ ,\qquad
		e^-_1 \zeta_1 \overline{\zeta^2} e^2_- \ .
	\end{split}
\end{equation}
Expanding $d(\zeta_1 \overline{\zeta^2})$ in this basis produces (\ref{eq:de1e2basis}). In a similar way, one can obtain a basis for odd forms; expanding $d(\zeta_1 \overline{\zeta_2})$ in this basis gives (\ref{eq:de1e2cbasis}).

$p^i$ and $q^i$ appear in (\ref{eq:de1e2basis}), (\ref{eq:de1e2cbasis}) as projected on different bases; for example, both $w_1 \cdot p_1$ and $w_2 \cdot p_1$ appear. Most of the projections are `mixed', i.e.~of the form $w_2 \cdot p_1$. The remaining, `unmixed' components can be eliminated by suitable linear combinations; for example, $e^-_1 \cdot {\rm Im} p_1$ can be obtained as the sum of the first parenthesis on the right hand side of (\ref{eq:de1e2basis}), and the first parenthesis on the right hand side of (\ref{eq:de1e2cbasis}). 

Unfortunately, one does not obtain all of the components of the $p^i$ and $q^i$ in this way. Looking again at the ten-dimensional system, we now get inspiration from (\ref{eq:LK}), and we look at the tensors
\begin{equation}
	\nabla_{(\mu} (k_1+k_2)_{\nu)} \ ,\qquad
	\nabla_{[\mu} (k_1-k_2)_{\nu]} \ .
\end{equation}
These are respectively the symmetrization and antisymmetrization of
\begin{equation}
	\nabla_\mu k_{1\,\nu} + \nabla_\nu k_{2\,\mu}\ .
\end{equation} 
We can compute this using
\begin{equation}\label{eq:nablaztorsi}
	\nabla_\mu k^i_\nu = 2 {\rm Re} (p^i_\mu k^i_\nu + 2 \bar q^i_\mu w^i_\nu)\ ,
\end{equation}
which is just the obvious extension of (\ref{eq:nablaztors}). The result can be expanded in components, obtaining (\ref{eq:nablazbasis}).

Together, (\ref{eq:de1e2basis}), (\ref{eq:de1e2cbasis}) and (\ref{eq:nablazbasis}) contain almost all of the components of $p^i$ and $q^i$. The only ones left are\footnote{This precisely parallels the fact that the intrinsic torsions $Q^1_{+_2 NP}$ and $Q^2_{+_1 NP}$ were the ones not determined by (\ref{eq:psp10}) and (\ref{eq:LK}); see \cite[(B.20)]{10d}.}
\begin{equation}\label{eq:rest}
	e^-_1\cdot p_2 \ ,\qquad e^-_2 \cdot p_1  \ ,\qquad
	e^-_1\cdot q_2 \ ,\qquad e^-_2 \cdot q_1  \ 
\end{equation}
(although, to be precise, the combination $e^-_1 \cdot {\rm Re} p_2 + e^-_2 \cdot {\rm Re} p_1$ does appear in (\ref{eq:nablazbasis})). 

To determine these remaining components using differential forms, we finally take inspiration from (\ref{eq:++1}), (\ref{eq:++2}): thus we look at $d(\zeta_1 \overline{\zeta^2} e^-_2)$, $d(\zeta_1 \overline{\zeta_2} e^-_2)$, $d(e^-_1 \zeta_1 \overline{\zeta^2})$, $d(e^-_1 \zeta_1 \overline{\zeta_2})$. In (\ref{eq:de-eps}) we have only kept the components that contain the torsions in (\ref{eq:rest}); the $\ldots$ terms are orthogonal to the ones shown, and will drop out from our considerations. For example, using (\ref{eq:de-eps}) and (\ref{eq:eeeC}) we can write
\begin{equation}\label{eq:eq}
	(d(e^-_1 \zeta_1 \overline{\zeta^2}), e^-_1 \zeta^1 \overline{\zeta^2}) = -16i e^-_1 \cdot \bar q_2 \ ,\qquad
	(d(\zeta_1 \overline{\zeta^2} e^-_2), \zeta_1 \overline{\zeta_2} e^-_2) = 16i e^-_2 \cdot q_1 \ .
\end{equation}
On the other hand, the components $e^-_1 \cdot p_2$ and $e^-_2 \cdot p_1$ appear in (\ref{eq:de-eps}) mixed with one-forms $\rho^i$ analogous to the $\rho$ appearing in (\ref{eq:dw}) and (\ref{eq:de-}). However, summing them appropriately and using (\ref{eq:dde-}) one obtains
\begin{equation}\label{eq:ep}
\begin{split}
	&(d(e^-_1 \zeta_1 \overline{\zeta_2}), e^-_1 \zeta^1 \overline{\zeta^2} e^-_2) +
	(d(e^-_1 \zeta^1 \overline{\zeta_2}), e^-_1 \zeta_1 \overline{\zeta^2} e^-_2) - 16i d^\dagger e^-_1 = 32i e^-_1 \cdot p_2\ , \\
	&(d(\zeta_1 \overline{\zeta^2} e^-_2), e^-_1 \zeta^1 \overline{\zeta_2} e^-_2)+
	(d(\zeta_1 \overline{\zeta_2} e^-_2), e^-_1 \zeta^1 \overline{\zeta^2} e^-_2) +16i d^\dagger e^-_2 = -32i e^-_2 \cdot p_1 \ .
\end{split}
\end{equation}
There is actually a small redundancy here, in that the real part of the sum of these two, $e^-_1 \cdot {\rm Re} p_2 + e^-_2 \cdot {\rm Re}  p_1$, is also determined by the $k^1_\mu k^2_\nu$ component of (\ref{eq:nablazbasis}).



\subsection{Gravitino} 
\label{sub:4dgravgen}


 
Our strategy will be to evaluate using supersymmetry the various intrinsic torsions we identified in section \ref{ssub:inttors2}. There are three classes of equations: those coming from $d(\zeta_1 \overline{\zeta^2})$, $d(\zeta_1 \overline{\zeta_2})$ (which we considered in (\ref{eq:de1e2})), those from $\nabla_\mu k_{1\,\nu} + \nabla_\nu k_{2\,\mu}$ (considered in (\ref{eq:nablaztorsi})), and the torsions ``along $k_i$'' ((\ref{eq:eq}), (\ref{eq:ep})). 

We can evaluate $d(\zeta_1 \overline{\zeta^2})$, $d(\zeta_1 \overline{\zeta_2})$ in the language of (\ref{eq:bisp12}). That means computing $d\mu$, $dv$, $d \omega$, $d*v$. Actually, before we give the expressions for these exterior derivatives, notice that one more exterior derivative is the antisymmetrization of $\nabla_\mu k_{1\,\nu} + \nabla_\nu k_{2\,\mu}$, namely $d(k_1-k_2)$. In (\ref{eq:timesusy}), we already gave formulas for $d(k_1-k_2)$ and $dv$, collected together in a triplet $d v_x$ (see (\ref{twisteddiff4d})). This time, however, we do not need to use $dk$. The exterior derivatives we need are then
\begin{subequations}\label{gravitinogeneral}
\begin{align}
	d\mu&= S_x v_x - \iota_K T^+\ , \\
	d v_x &= 2\epsilon_{xyz} {\rm Im} ( \bar S_y \omega_z )\ , \\
	d \omega &= i \hat Q \wedge \omega +\frac i2 (A^+ \wedge \omega_1 - A^- \wedge \omega_2)+\frac 32 i * ( - S_+ \bar v +  S_-  v + S_3 (k_1+k_2)) - (k_1 - k_2 )\wedge T^+ \ , \\
	d*v &= -2 (\bar \mu S^+ + \mu \overline{S^-}) \vol_4 \ .
\end{align}	
\end{subequations}

We also have to compute the symmetrization of $\nabla_\mu k_{1\,\nu} + \nabla_\nu k_{2\,\mu}$, which simply gives
\begin{equation}\label{eq:killing4}
	\nabla_{(\mu} k_{\nu)} = 0 \ .
\end{equation}
This is simply the statement that $k$ is a Killing vector. 

Finally, we have to compute the remaining torsions (\ref{eq:eq}), (\ref{eq:ep}). These give
\begin{subequations}\label{eq:e-gen}
\begin{align}
	&e^-_1 \cdot (4 q_2 + S^- k_2 +S^3 v  -2 \iota_v T^+ - i \mu A^-)=0\ ,\\
	&e^{-\,\mu}_1  \left( 2 p_{2\,\mu} -  
	S^- \bar w_2^\mu
	- (\bar \lambda e^-_2 - \iota_{e^-_2} \bar \omega)^\nu \left(T^+_{\mu\nu} - \frac12 S^3 g_{\mu\nu}\right)-2i \hat Q_\mu -i e^-_2 \cdot v A^-_\mu\right) =0\ ,\\
	&e^-_2 \cdot (4 q_1 - S^+ k_1 +S^3 \bar v  +2 \iota_{\bar v} T^+ 
	+i \mu A^+)=0\ ,\\
	&e^{-\,\mu}_2  \left( 2 p_{1\,\mu} +  
	S^+ \bar w_1^\mu
	- (\bar \lambda e^-_1 + \iota_{e^-_1 } \bar \omega)^\nu \left(T^+_{\mu\nu} + \frac12 S^3 g_{\mu\nu}\right) -2i \hat Q_\mu -i e^-_1 \cdot \bar v A^-_\mu \right) =0 \ .
\end{align}
\end{subequations}
(Recall that $e^- \cdot q$ and $e^- \cdot p$ are given in terms of forms by (\ref{eq:eq}), (\ref{eq:ep}).) As remarked after (\ref{eq:ep}), there is actually a slight redundancy: $e^-_1 \cdot {\rm Re} p_2 + e^-_2 \cdot {\rm Re}  p_1$ is also determined by the $k^1_\mu k^2_\nu$ component of (\ref{eq:nablazbasis}). 


\subsection{Hyperino equations} 
\label{sub:4dhynogen}

Now we turn to the hyperino equations (\ref{eq:4dhyno}). As we remarked in section \ref{sub:4dhyno} to extract all the information contained in these equations it is sufficient to expand them in basis of spinors. Moreover, as explained in section \ref{ssub:inttors2} two possible bases of spinors of positive chirality are given by
\begin{align}
\label{eq:basisspinors}
 & \zeta_i \ ,  \quad e_i^- \zeta^i\ , \qquad i= 1,\,2 \ .
\end{align}

In principle a minimal set of equations which is equivalent to the original hyperino equations can be obtained by choosing a specific basis (for example the bases with index $i=1$) and expanding the (\ref{eq:4dhyno}) along this basis. However, it turns out that it is more convenient to expand (\ref{eq:4dhyno}) along {\it all} the spinors appearing in (\ref{eq:basisspinors}). Of course the resultant system of equations will be redundant but it can be written in a neat form.

Obviously, when we expand (\ref{eq:4dhyno}) along $\zeta_i$ we simply re-obtain equations (\ref{eq:timesusyhyper}). However, in this case these equations are not sufficient, and they must be completed by the equations obtained by expanding (\ref{eq:4dhyno}) along $e_i^- \zeta^i$.
To this end, it is useful to introduce
\begin{equation}\label{eq:bilinearsgen1}
    E^{k i}_\mu \equiv 
	\bar{\zeta}^k e_k^- \gamma_\mu \zeta^i=
	-4\left(
	\begin{array}{cc}
	2 \bar{w}_1 & \bar \mu e^-_1 + \iota_{e^-_1} \bar\omega \\
	- \bar \mu e^-_2 + \iota_{e^-_2} \bar \omega & 	2 \bar{w}_2 
	\end{array}
	\right)_\mu = E_\mu \epsilon^{k i} + E_{\mu\,x} \sigma^{x ik} \ , 
\end{equation}
where $k$ is not summed. This matrix does not actually transform well under SU(2); it is a bookkeeping device. It would be possible to define a three-index tensor $\bar{\zeta}^k e_j^- \gamma^\mu \zeta^i$ that does transform well, but for our current aim this would be an overkill. We also introduce
\begin{align}\label{eq:SU(2)bilinearsgen2}
  &  C^{k}{}_i\equiv\bar{\zeta}^k e_k^- \zeta_j = 
	4 \left( \begin{array}{cc}
		2 & e_1^- \cdot \bar{v}\\ e_2^- \cdot v & 2
	\end{array}\right) =  C \delta ^{k}{}_i + C^x \sigma^{x\,k}{}_i\, .
\end{align}

Returning now to (\ref{eq:4dhyno}) we can expand them along $\bar{\zeta}^k e_k^-$ and, after some manipulations very similar to those that lead to (\ref{eq:timesusyhyper}), we obtain the equation
\begin{equation}
\label{eq:gensusyhyper}
i E \cdot  D q^v - \Omega^{x\,v}{}_u E^x \cdot D q^u - C g \bar{\mathcal{L}}^{\Lambda} k_{\Lambda}^v - i g \Omega^{x\, v}{}_u C^x \bar{\mathcal{L}}^{\Lambda} k_{\Lambda}^{u} = 0\ .
\end{equation}
In the general case, the hyperino equations are equivalent, in a slightly redundant manner, to (\ref{eq:gensusyhyper}) and (\ref{eq:timesusyhyper}).



\subsection{Gaugino equations} 
\label{sub:4dhgauginogen}

Finally we move to the gaugino equations (\ref{eq:4dgaugino}). Contrary to the hyperino equations just discussed, in this case it is not convenient to use the SU(2) formalism; therefore we will rewrite (\ref{eq:4dgaugino}) as
\begin{subequations}\label{eq:gauginogen}
	\begin{align}
	\label{eq:gauginogen1}
	  & i\, D t^a \zeta^i + G^{a\,+} \zeta_2 + W^a \zeta_2 - \frac{i}{2} W^{a-} \zeta_1 + \frac{i}{2} W^{a 3} \zeta_2 =0 \ , \\
	\label{eq:gauginogen2}
	  & i\, D t^a \zeta^2 - G^{a\,+} \zeta_1 - W^a \zeta_1 + \frac{i}{2} W^{a 3} \zeta_1 + \frac{i}{2} W^{a+} \zeta_2 = 0 \ .
	\end{align}	
\end{subequations}
To proceed we expand equations (\ref{eq:gauginogen1}) along $\bar{\zeta}_2$ and $\bar{\zeta}^2 e_2^-$, and (\ref{eq:gauginogen2}) along $\bar{\zeta}_1$ and $\bar{\zeta}^1 e_1^-$ obtaining the system
\begin{subequations}\label{eq:gensusygauginototal}
\begin{align}
		&i\,\bar{v} \cdot D t^a + G^{a +} \llcorner \omega_2 - \frac{i}{2} \mu W^{a -} =0 \ ,\\
		&- i \bigl((\iota_{e_2^-} \bar{\omega}) - \bar{\mu} e_2^- \bigr) \llcorner D t^a - 2 \iota_{e_2^-} \iota_{k_2} G^{a +} + 2 W^a - \frac{i}{2} (e_2^- \cdot v) W^{a -} + i W^{a 3} =0\ , \\
		&i\,v \cdot D t^a - G^{a +} \llcorner \omega_1 - \frac{i}{2} \mu W^{a +} =0 \ ,\\
		&- i \bigl((\iota_{e_1^-} \bar{\omega}) + \bar{\mu} e_1^- \bigr) \llcorner D t^a + 2 \iota_{e_1^-} \iota_{k_1} G^{a +} - 2 W^a + \frac{i}{2} (e_1^- \cdot \bar{v}) W^{a +} + i W^{a 3} =0
		\ .
\end{align}
\end{subequations}
The system (\ref{eq:gensusygauginototal}) is therefore completely equivalent to the original equation (\ref{eq:4dgaugino}) without any redundancy.


\subsection{Summary: four-dimensional general case} 
\label{sub:4dgensum}

For four-dimensional gauged ${\cal N}=2$ supergravity, in the general case, preserved supersymmetry is equivalent to the system given by (\ref{gravitinogeneral}), (\ref{eq:killing4}), (\ref{eq:e-gen}), (\ref{eq:gensusyhyper}), (\ref{eq:timesusyhyper}), (\ref{eq:gensusygauginototal}). As anticipated, this system is less pleasant-looking than the ``boxed'' system in section \ref{sec:4dsugratime}, which is appropriate for the timelike case. 



\section{Conclusions} 
\label{sec:conc}
Let us summarize here our main results. We have written the conditions for preserved supersymmetry in ten dimensional type II supergravity, for a (topologically trivial) fibration, in a way that parallels the supersymmetry conditions in gauged ${\cal N}=2$ supergravity in four dimensions. 

In section \ref{sec:4dsugratime} we gave a minimal (see footnote \ref{foot:MO}) system which is equivalent to preserved supersymmetry for gauged four-dimensional ${\cal N}=2$ supergravity:
\begin{subequations}
\begin{align}
	&\label{eq:mu4}
	D\mu= S_x v_x - 2\iota_k T^+\ ,\\
			& d k = -2 {\rm Re} ( S_x \bar o_x  + 2 \bar\lambda T^+ )\ ,\\
			& D v_x = 2\epsilon_{xyz} {\rm Im} ( \bar S_y  o_z ) \ ; \\
	& \label{eq:hm4} i\, k \cdot D q^v + \Omega^{xv}{}_u v_x \cdot D q^u - g\,\bar{\mathcal{L}}^{\Lambda}k_{\Lambda}^{v} \, \mu = 0\ ; \\
	& \label{eq:vm4} 2 i \bar{\mu} D t^a - 4 \iota_k G^{a +} - i W^{a x} v^x + 2 W^a k = 0\ .
\end{align}	
\end{subequations}
The first three equations come from the gravitino; (\ref{eq:hm4}) from the hyperinos, and (\ref{eq:vm4}) from gauginos.

In section \ref{sec:time10d} we gave a minimal system equivalent to preserved supersymmetry in the timelike case. We consider a spinor Ansatz (\ref{eq:spAn}), (\ref{eq:eqnorms}); our assumptions on the metric and on the $B$ field are summarized in footnote \ref{foot:fibr} and in (\ref{eq:H1H20}). 
The conditions for unbroken supersymmetry can be divided in three different families. First we have the equations for the external gravitino:
\begin{subequations}\label{eq:summary10dtime}
\begin{align}
  &(\ref{eq:mu4})  \ {\rm with} \ (\ref{twisteddiff10}) \ , \label{eq:dmu10conc}\\ 
	&d_4 v^3 = \iota_k H_3 \ , \label{eq:dv^310conc}\\ 	
  & d_4 v + 2i v \wedge  (\bar \phi_- , e^\phi d_4 (e^{-\phi} \phi_-)) = - \frac 12 \omega \bar{s}_- + \frac 12 \bar{\omega} s_+ - i e^\phi \iota_k (\bar{\phi}_-, F_3) - i e^\phi v^3 \wedge (\bar{\phi}_-, F_1) \ , \label{eq:dv10conc}\\
	&d_4 k - \iota_{v^3} H  = 2 \mathrm{Re} \left[-  \bar{\omega} \,s_3 + e^\phi(\bar{\phi}_-, F_1) \llcorner (\ast \bar{v}) - 2\bar{\mu}\, T^+ \right] \label{eq:dk10conc} \ . 
\end{align}
(\ref{eq:dv^310conc}), (\ref{eq:dv10conc}) and (\ref{eq:dk10conc}) are not manifestly equivalent to the corresponding four-dimensional ones. It would be possible to make them more similar by using the equations for the universal hypermultiplet (\ref{eq:univ}). However, even so they would not become identical: this is because we are working in the string frame, and without redefining the four-dimensional metric to go to the Einstein frame (as we already remarked in section \ref{sub:LK}). We then preferred keeping the expression we gave, because for example (\ref{eq:dv^310conc}) is simpler this way. 

The second set of equations corresponds to multiplets which are loosely speaking related to forms in the ``bulk'' of the Hodge diamond (\ref{eq:hodge}): these are the multiplets that can be included in a reduction to effective ${\cal N}=2$ supergravity \cite{grana-louis-waldram}. They include the universal hypermultiplet, the non-universal hypermultiplet and by the ``bulk'' vector multiplet: 
\begin{align}	
	&(\ref{eq:hm4})\ {\rm with}\ (\ref{eq:uhyper}), (\ref{eq:Ki10}), (\ref{eq:gaugvu})\ ; \\
	&(\ref{eq:hm4})\ {\rm with}\ (\ref{eq:nuhyper}), (\ref{eq:Ki10}), (\ref{eq:gaugvnu})\ ; \\
	&(\ref{eq:vm4}) \ {\rm with}\ (\ref{eq:Dt}), (\ref{eq:G+}) \ .    
\end{align}	
In black hole applications, these first-order equations would become the attractor equations for hypermultiplet scalars (which very often do not flow at all) and for vector multiplet scalars. 

Finally, we have equations corresponding to forms coming from the ``edge'' of the diamond (\ref{eq:hodge}): these would correspond a new ``edge'' vector multiplet, and a gravitino multiplet. These cannot be easily included in a reduction approach, because extra gravitinos would be inconsistent without extra supersymmetry; but they do appear in our approach, in which we are not reducing but simply reorganizing the ten-dimensional equations. The equations read, for the ``edge'' vector multiplet:
\begin{align}\label{vmedge}
\mu (\gamma^{\bar{i}_1} \phi_+, F_1) = 2 e^{-\phi} (\gamma^{\bar{i}_1} \phi_+, d_6 \bar{\phi}_+) (k - v^3) + (\gamma^{\bar{i}_1} \phi_-, F_0) v - 2 \left(\frac{e^{-\phi}}{\bar{\mu}} v \cdot (\gamma^{\bar{i}_1} \phi_-, d_4 \phi_+) \right) v^3\ ,\\
	\bar\mu (\bar\phi_+ \gamma^{\bar i_2}, F_1) = 2 e^{-\phi} (\bar\phi_+ \gamma^{\bar i_2}, d_6 \phi_+) (k + v^3) + (\phi_- \gamma^{\bar i_2}, F_0) v + 2 \left(\frac{e^{-\phi}}{\mu} v \cdot (\bar\phi_+ \gamma^{\bar i_2}, d_4 \phi_-) \right) v^3\ ;
\end{align} 
and, for the gravitino multiplets:
\begin{align}\label{gravitinomultconc}
&\iota_k (\gamma^{i_1} \bar\phi_-, F_2) = - 2 \bar\mu e^{-\phi} (\gamma^{i_1} \bar\phi_-, d_4 \bar\phi_+) - 2 Q^x_L v_x \ , 	\\
	 &\iota_k (\phi_- \gamma^{\bar i_2}, F_2) = - 2 \bar\mu e^{-\phi} (\phi_- \gamma^{\bar i_2}, d_4 \bar\phi_+) - 2 Q^x_R v_x \ , \\
	&\iota_{v^3} (\gamma^{\bar{i}_1}\phi_-, d_4 \phi_+) = \iota_{v^3} (\phi_- \gamma^{\bar i_2}, d_4 \bar\phi_+) = 0 = \iota_{k} (\gamma^{\bar{i}_1}\phi_-, d_4 \phi_+) = \iota_{k} (\phi_- \gamma^{\bar i_2}, d_4 \bar\phi_+)  \ , \\
	 & (\gamma^{\bar{i}_1} \phi_+, d_6 \bar{\phi}_+) = (\gamma^{\bar{i}_1} \phi_-, d_6 \bar{\phi}_-)
	\ ,\qquad
	 (\bar\phi_+ \gamma^{\bar i_2}, d_6 \phi_+) =- (\phi_- \gamma^{\bar i_2}, d_6 \bar{\phi}_-) \ ,
\end{align}
\end{subequations}
where $Q^x_L$ and $Q^x_R$ are given by (\ref{Q^x}). 

Together, equations (\ref{eq:summary10dtime}) express the conditions for unbroken supersymmetry in the ten-dimensional timelike case.

We also dealt with the general case, where the four-dimensional spinors can coincide on some locus in spacetime. This case is non-generic, but it is still important for ${\cal N}=1$ vacua and for domain walls. Our systems for those cases are summarized in sections \ref{sub:10dsum} and \ref{sub:4dgensum}; they are much more involved than the systems we just summarized for the timelike case, and they will not be repeated here.


\section*{Acknowledgments}
We would like to thank L.~Andrianopoli, M.~Gra\~na, K.~Hristov, P.~Meessen, A.~Sagnotti and A.~Zaffaroni for interesting discussions. 
The authors are supported in part by INFN, by the MIUR-FIRB grant RBFR10QS5J ``String Theory and Fundamental Interactions'', and by the MIUR-PRIN contract 2009-KHZKRX. The research of A.T.~is also supported by the ERC Starting Grant 307286 (XD-STRING).

\appendix

\section{Bilinears} 
\label{sec:bil}

The usual Chevalley--Mukai pairing on forms reads
\begin{equation}\label{eq:mukai}
	( \alpha , \beta )\equiv (\alpha \wedge \lambda \beta)_{\rm top}
\end{equation}
where  $\lambda$ was defined in (\ref{eq:gl}), and ${}_{\rm top}$ means keeping the $d$-form part, and dividing by the volume form. The pairing (\ref{eq:mukai}) is related to the trace of bispinors by 
\begin{equation}\label{eq:mukaitr}
	(A,B)= (-)^{{\rm deg}(A)} \frac1{2^{\lfloor \frac d2 \rfloor}} {\rm Tr}( *A \, B) \ .
\end{equation}

When either $A$ or $B$ in (\ref{eq:mukaitr}) is defined in terms of chiral spinors, we can simplify the formula further by computing the action of the Hodge star. Let us consider positive chirality spinors $\zeta_i$ in four dimensions, as in section \ref{sub:4d2eps}.  Using (\ref{eq:gdagger}), we can also compute 
\begin{equation}\label{eq:e2e1}
	\zeta_2 \otimes \overline{\zeta^1} = \lambda(\zeta^1 \otimes \overline{\zeta_2}) \ ,\qquad
	\zeta_2 \otimes \overline{\zeta_1} = - \lambda(\zeta_1 \otimes \overline{\zeta_2})\ .
\end{equation}
So for example we can compute, using (\ref{eq:gl}) and (\ref{eq:e2e1}):
\begin{equation}
	* \zeta_1 \overline{\zeta^2} = - i \gamma \lambda(\zeta_1 \overline{\zeta^2}) = i \zeta^2 \overline{\zeta_1}\ . 
\end{equation}
From this we get
\begin{equation}\label{eq:e1e2C}
	(\zeta_1 \overline{\zeta^2},C) = - \frac14{\rm Tr} ( *(\zeta_1 \overline{\zeta^2}) C) = -\frac i4 \overline{\zeta_1} C \zeta^2\ .
\end{equation}
In a similar way, one can find
\begin{equation}\label{eq:e1e2cC}
	(\zeta_1 \overline{\zeta_2}, C) = -\frac i4 \overline{\zeta_1} C \zeta_2 \ .
\end{equation}
A little more work, and the formula $\lambda(\gamma^M C) = \lambda(C) \gamma^M$ (see \cite[(A.6)]{10d}), shows
\begin{align}\label{eq:eeeC}
\begin{split}
	&(e^-_1 \zeta_1 \overline{\zeta^2}, C) = \frac i4 \overline{\zeta_1} e^-_1 C \zeta^2 \ ,\qquad
	(e^-_1 \zeta_1 \overline{\zeta_2}, C) = -\frac i4 \overline{\zeta_1} e^-_1 C \zeta_2 \ ;\\
	&(\zeta_1 \overline{\zeta^2}e^-_2, C) = -\frac i4 \overline{\zeta_1} C e^-_2 \zeta^2 \ ,\qquad
	(\zeta_1 \overline{\zeta_2}e^-_2, C) = \frac i4 \overline{\zeta_1} C e^-_2 \zeta_2 \ ;\\
	&(e^-_1 \zeta_1 \overline{\zeta^2} e^-_2,C)= \frac i4
	\overline{\zeta_1} e^-_1 C e^-_2 \zeta^2 \ ,\qquad
	(e^-_1 \zeta_1 \overline{\zeta_2} e^-_2, C)= -\frac i4
	\overline{\zeta_1} e^-_1 C e^-_2 \zeta_2\ .
\end{split}
\end{align}


\section{Bispinor reduction} 
\label{sec:Phi46}

In the main text, we often need to translate a form obtained as a bispinor in ten dimensions to a wedge of bispinors in four and six dimensions. While the reduction is conceptually straightforward, it gives rise to some subtle signs. 

We will show here, as an example, how to reduce $\Phi= \epsilon_1 \overline{\epsilon_2}$ using our decomposition (\ref{eq:spAn}). Using (\ref{eq:Gamma46}), one computes
\begin{align}
\begin{split}
	\epsilon_1 \overline{\epsilon_2} &= \sum \frac1{32 i!} \overline{\epsilon_2} \Gamma_{M_i \ldots M_1} \epsilon_1 dx^{M_1}\wedge \ldots \wedge dx^{M_i} \\
	& = \sum \frac1{32 k! j!} \overline{\epsilon_2} \Gamma_{m_j \ldots m_1 \mu_i \ldots \mu_i} \epsilon_1 dx^{\mu_1} \wedge \ldots \wedge dx^{\mu_k}\wedge dy^{m_1}\wedge \ldots \wedge dy^{m_j} \\
	& = \sum \frac1{32 k! j!} \Big[ \overline{\zeta_\mp} \gamma_5^j \gamma_{\mu_k \ldots \mu_1 } \zeta_+ \eta^{2\,\dagger} \gamma_{m_j\ldots m_1} \eta^1_+ + 
	\overline{\zeta_\mp} \gamma_5^j \gamma_{\mu_k \ldots \mu_1 } \zeta_- \eta^{2\,\dagger} \gamma_{m_j\ldots m_1} \eta^1_- + {\rm c.c.} 
	\Big]\\
	& = \zeta_+ \overline{\zeta_\mp} \wedge \eta^1_+ \eta^{2\,\dagger}_+ 
	\pm \zeta_- \overline{\zeta_\mp} \wedge \eta^1_- \eta^{2\,\dagger}_+ 
	\mp \zeta_+ \overline{\zeta_\pm} \wedge \eta^1_+ \eta^{2\,\dagger}_-
	+ \zeta_- \overline{\zeta_\pm} \wedge \eta^1_- \eta^{2\,\dagger}_-
	 \ .
\end{split}
\end{align}
All bispinors in the last line of this formula should be understood as forms: for example, $\eta^1_+ \eta^{2\,\dagger}_+= \sum \frac1{8j!} \eta^{2\,\dagger} \gamma_{m_j\ldots m_1} \eta^1_+ dy^{m_1}\wedge \ldots \wedge dy^{m_j}$. One subtlety is that $\eta^1_- \eta^{2\,\dagger}_+= - \overline{\phi_-}$, which comes about because the gamma matrices in six dimensions are purely imaginary. Taking this into account, one obtains (\ref{eq:Phi46}). 


\section{Generalized structures in four dimensions} 
\label{app:pure}

We have seen in section \ref{sub:4d2eps} that the structure defined by the two spinors $\zeta_i$ is $\rr^2$ in the null case (where the $\zeta_i$ are proportional) and the identity group in the timelike case (where the $\zeta_i$ are not proportional). We will now notice that their ``generalized structure'', namely the structure they define on $T \oplus T^*$, is always the same. 

This has to do with the fact that the $\zeta_i$ are \emph{pure spinors}: namely, their annihilator in ${\rm Cl}(1,3)$ is half the dimension of the spacetime, in this case two. For each of the $\zeta_i$, we saw this in (\ref{eq:4dpure}). In Euclidean signature the stabilizer of a pure spinor would be ${\rm SU}(2)$; in Lorentzian signature, as we have seen in section \ref{sub:4deps}, the stabilizer of a four-dimensional spinor is $\rr^2$. 

Let us now consider the bispinors (\ref{eq:bisp12}). The structure they define on $T \oplus T^*$ is the stabilizer in Spin$(4,4)$, the ``generalized Lorentz group'' that acts on it. This is generated by the action of operators of the form 
\begin{equation}\label{eq:infgenSp}
	\omega_{AB} \Gamma^{AB}
\end{equation}
where the $\Gamma^A$ consist of the usual gamma matrices acting from the left and from the right on the bispinor, which we will denote by $\stackrel\to \gamma_\mu$ and $\stackrel \leftarrow \gamma_\mu$ respectively. Under the Clifford map that maps bispinors to forms (see footnote \ref{foot:clifford}), these gamma matrices are related to wedges and contractions: 
\begin{equation}
	\label{eq:gwedge}
		\gamma^\mu  C_k = (dx^\mu\wedge + \iota^\mu)  C_k
		\ ,\qquad
		C_k \gamma^\mu = (-)^k (dx^\mu \wedge - \iota^\mu) C_k\ .
\end{equation}
So another basis for the gamma matrices $\Gamma^A$ in (\ref{eq:infgenSp}) is given by operators $dx^\mu$ and $ \iota_\mu$, acting on forms. Their algebra is that of a Clifford algebra with respect to the metric ${\cal I}= \left(\begin{smallmatrix} 0 & 1 \\ 1 & 0 \end{smallmatrix}\right)$, which is nothing but the pairing between one-forms and vectors. Since this metric has signature $(4,4)$ (even when the spacetime has Lorentzian signature), this Clifford algebra is called ${\rm Cl}(4,4)$. (For more details about generalized structure group, see for example \cite{gualtieri}, \cite[Sec.~3]{gmpt3}, \cite[Sec.~2]{10d}.) 

Computing the stabilizer of a form in ${\rm Sp}(4,4)$ is often easier in the bispinor picture. Let us start by considering one of the bispinors in (\ref{eq:bisp12}), for example $\zeta_1 \otimes \overline{\zeta_2}$. Its annihilator consists of $k_1$ and $w_1$ acting from the left, and of $k_2$ and $\bar w_2$ acting from the right:
\begin{equation}
	{\rm Ann}(\zeta_1 \otimes \overline{\zeta^2}) = \{ \stackrel \to k_1 \, , \stackrel \to w_1 \, , \stackrel \leftarrow k_2 \, , \stackrel \leftarrow {\bar w_2} \}\ .
\end{equation}
Thus $\zeta_1 \otimes \overline{\zeta^2}$ is a pure spinor for ${\rm Cl}(4,4)$. The stabilizer of such a pure spinor is ${\rm SU}(2,2)$. A similar discussion holds for $\zeta_1 \otimes \overline{\zeta_2}$. 

We finally compute the common stabilizer of $\zeta_1 \otimes \overline{\zeta^2}$ and $\zeta_1 \otimes \overline{\zeta_2}$. In four Euclidean dimensions, the common stabilizer of two pure spinors of this form would be ${\rm SU}(2)\times {\rm SU}(2)$. In Lorentzian signature, however, this is no longer the case. Explicitly, we find the generators
\begin{equation}
	{\rm Stab}(\zeta_1 \otimes \overline{\zeta^2},\zeta_1 \otimes \overline{\zeta_2}) = {\rm span}\left\{
	\begin{array}{c}
	\stackrel \to \gamma_{-_1 a_1} \, ,	\
	\stackrel \to \gamma_{-_1} \stackrel \leftarrow \gamma_{a_2}\, , \
	\stackrel \leftarrow \gamma_{-_2 a_2} \, , \
	\stackrel \to \gamma_{a_1} \stackrel \leftarrow \gamma_{-_2} \, , \
	\stackrel \to \gamma_{-_1} \stackrel \leftarrow \gamma_{-_2}
	 \vspace{.1cm}\\
	\stackrel \to \gamma_{+_1 -_1} + \stackrel \leftarrow \gamma_{+_2 -_2}\, , \
	\stackrel \to \gamma_{+_1} \stackrel \leftarrow \gamma_{-_2}\, , \ 
	\stackrel \to \gamma_{-_1} \stackrel \leftarrow \gamma_{+_2}
	\end{array}
 \right\}\ .
\end{equation}
Here we think of $k_i$ as $e_{-_i}$, while the index $a_i$ runs over $w_i$ and $\bar w_i$. The Lie algebra structure can now be computed similarly as in \cite[Sec.~2.2.1]{10d}. The generators of the first line generate a Heisenberg algebra with nine generators, ${\rm Heis}_9$; the second line generates a copy of ${\rm Sl}(2,\rr)$. Moreover, the second subalgebra acts on the first, so that the total algebra is that of a semidirect product: 
\begin{equation}\label{eq:gsLie}
	{\rm Sl}(2,\rr) \ltimes {\rm Heis}_9 \qquad ({\rm on} \ T \oplus T^*)\ .
\end{equation}

Thus the forms in (\ref{eq:bisp12}) define on $T \oplus T^*$ the structure given by (\ref{eq:gsLie}), both in the timelike and null case.  The precise form of the algebra (\ref{eq:gsLie}), however, will actually have no role in our paper. 


\section{Details about four-dimensional intrinsic torsions} 
\label{sec:diff}

\subsection{Expansion in components} 
\label{sub:expdiff}

We will evaluate here the derivatives we need in our discussion in section \ref{sub:torsgen}. First we evaluate the exterior derivatives $d(\zeta_1 \overline{\zeta^2})$ and $d(\zeta_1 \overline{\zeta_2})$, expanded in the basis (\ref{eq:evenforms}) of even forms, and in the analogue basis of odd forms:
\begin{align}\label{eq:de1e2basis}
	\begin{split}
		d (\zeta_1 \overline{\zeta^2}) = &
		 \Big(-\frac12w_1 \cdot (p_1 + \bar p_2) + e_1^-\cdot q_1\Big)\zeta^1 \overline{\zeta^2}
		+\Big(\frac14 k_1 \cdot (p_1 + \bar p_2) + \frac12 \bar w_1 \cdot q_1 \Big) e^1_- \zeta_1 \overline{\zeta^2}\\
		+&\frac12 (w_1\cdot \bar q_2) \zeta^1 \overline{\zeta_2} e^-_2
		-\frac14 (k_1\cdot \bar q_2) e^-_1 \zeta_1 \overline{\zeta_2} e^-_2\\
		+ &\Big(- \frac12 \bar w_2 \cdot (p_1 + \bar p_2 )+e^-_2 \cdot \bar q_2\Big)\zeta_1 \overline{\zeta_2}
		-  \frac12 (\bar w_2 \cdot q_1) e^-_1 \zeta^1 \overline{\zeta_2} \\
		-&  \Big(\frac14 k_2 \cdot (p_1 + \bar p_2) +\frac12 w_2 \cdot \bar q_2 \Big)\zeta_1 \overline{\zeta^2}e^-_2
		-\frac14 k_2 \cdot q_1 e^-_1 \zeta^1 \overline{\zeta^2} e^-_2\ .
	\end{split}
\end{align}
\begin{align}\label{eq:de1e2cbasis}
	\begin{split}
		d (\zeta_1 \overline{\zeta_2}) = &
		 \Big(-\frac12 w_1 \cdot (p_1 + p_2) + e_1^-\cdot q_1\Big)\zeta^1 \overline{\zeta_2}
		+\Big(\frac14 k_1 \cdot (p_1 + p_2) + \frac12 \bar w_1 \cdot q_1 \Big) e^1_- \zeta_1 \overline{\zeta_2}\\
		+&\frac12 (w_1\cdot \bar q_2) \zeta^1 \overline{\zeta^2}e^-_2 
		-\frac14 (k_1\cdot \bar q_2) e^-_1 \zeta_1 \overline{\zeta^2} e^-_2\\
		+ &\Big(\frac12 w_2 \cdot (p_1 + p_2 )-e^-_2 \cdot \bar q_2\Big)\zeta_1 \overline{\zeta^2}
		+  \frac12 ( w_2 \cdot q_1) e^-_1 \zeta^1 \overline{\zeta^2} \\
		+&  \Big(\frac14 k_2 \cdot (p_1 + \bar p_2) +\frac12 w_2 \cdot \bar q_2 \Big)\zeta_1 \overline{\zeta_2}e^-_2
		+\frac14 k_2 \cdot q_1 e^-_1 \zeta^1 \overline{\zeta_2} e^-_2\ .
	\end{split}
\end{align}
The next item we need is $\nabla_\mu k^1_\nu + \nabla_\nu k^2_\mu$. This is a rank 2 tensor, neither symmetric nor antisymmetric. We expanded in a basis obtained by tensoring the basis $\{ k^1, e^1_-, w^1, \bar w^1 \}$ with the similar basis $\{ k^2, e^2_-, w^2, \bar w^2 \}$. We obtain  
\begin{align}\label{eq:nablazbasis}
		&\nabla_\mu k^1_\nu + \nabla_\nu k^2_\mu=\\
		\nonumber &k^1_\mu k^2_\nu \Big( e^1_- \cdot {\rm Re} p_2 + e^2_- \cdot {\rm Re} p_1 \Big) + e^1_{-\,\mu} k^2_\nu \Big( k^1\cdot {\rm Re} p_2 \Big) + \left[w^1_\mu k^2_\nu \Big( \bar w^1 \cdot {\rm Re} p_2 + e^2_- \cdot \bar q_1 \Big) + {\rm c.c.}\right]\\
		\nonumber +&k^1_\mu e^2_{-\,\nu} \Big( k^2\cdot {\rm Re} p_1\Big) \hspace{5.6cm}  +\left[w^1_\mu e^2_{-\,\nu} \Big( k^2\cdot \bar q_1\Big) + {\rm c.c.} \right]\\
		\nonumber +& \left[ k^1_\mu w^2_\nu \Big( \bar w^2 \cdot {\rm Re} p_1 + e^1_- \cdot \bar q_2\Big) + e^1_{-\,\mu} w^2_\nu \Big(k^1\cdot \bar q_2\Big) + w^1_\mu w^2_\nu \Big( \bar w_2 \cdot \bar q_1 + \bar w_1 \cdot \bar q_2\Big) + \bar w^1_\mu w^2_\nu \Big( \bar w_2 \cdot q_1 + w_1 \cdot \bar q_2\Big) \right.\\
		\nonumber +&\left. {\rm c.c.} \right]\ .
\end{align}
We finally need the exterior differentials $d(\zeta_1 \overline{\zeta^2} e^-_2)$, $d(\zeta_1 \overline{\zeta_2} e^-_2)$, $d(e^-_1 \zeta_1 \overline{\zeta^2})$, $d(e^-_1 \zeta_1 \overline{\zeta_2})$, just like we did for (\ref{eq:de1e2basis}), (\ref{eq:de1e2cbasis}):
\begin{subequations}\label{eq:de-eps}
	\begin{align}
		d(e^-_1 \zeta_1 \overline{\zeta^2}) &= \Big( e^-_1 \cdot (- \bar p_1 + \bar p_2) + 2 \bar w_1 \cdot \rho_1 \Big) \zeta_1 \overline{\zeta^2} - \Big( e^-_1 \cdot \bar q_2 \Big) \zeta_1 \overline{\zeta_2} e^-_2 + \ldots \ , \\
		d(e^-_1 \zeta_1 \overline{\zeta_2}) &= \Big( e^-_1 \cdot (- \bar p_1 +  p_2) + 2 \bar w_1 \cdot \rho_1 \Big) \zeta_1 \overline{\zeta_2} - \Big( e^-_1 \cdot \bar q_2 \Big) \zeta_1 \overline{\zeta^2} e^-_2 + \ldots \ ; \\
		d(\zeta_1 \overline{\zeta^2} e^-_2) &= \Big( e^-_2 \cdot (p_1 - p_2) + 2 w_2 \cdot \bar \rho_2 \Big) \zeta_1 \overline{\zeta^2} + \Big(  e^-_2 \cdot q_1 \Big) e^-_1 \zeta^c_1 \overline{\zeta^2}  + \ldots \ ,\\
		d(\zeta_1 \overline{\zeta_2} e^-_2) &= \Big( -e^-_2 \cdot (p_1 - \bar p_2) - 2 \bar w_2 \cdot \rho_2 \Big) \zeta_1 \overline{\zeta_2} - \Big( e^-_2 \cdot q_1 \Big) e^-_1 \zeta^c_1 \overline{\zeta_2}  + \ldots \ . 		
	\end{align}	
\end{subequations}
In this case we did not give the whole expression; we only kept the components which are relevant for our needs. 

\subsection{Necessity of pairing equations} 
\label{sub:necpairing}

As we have seen, in four dimensions the gravitino equations (\ref{eq:timesusy}) are much easier than their counterpart (\ref{gravitinogeneral}) for the general case. This simplification depends in part on having added the exterior differential $dk$ to the system. In the original ten-dimensional system (\ref{eq:susy10}), there is no equation for $dK$. One might therefore wonder whether in the timelike case the pairing equations (\ref{eq:++1}), (\ref{eq:++2}) might be dropped out, and equations (\ref{eq:LK}) and (\ref{eq:psp10}) are sufficient to fully reconstruct the four-dimensional intrinsic torsions.

To this end we have to consider equations (\ref{eq:de1e2basis}), (\ref{eq:de1e2cbasis}) and (\ref{eq:nablazbasis}). Let us recall the identifications\footnote{In this appendix, we systematically drop the index $1$ from vectors and torsions. We also remove dependence from $\mu$ since this is completely determined from the equation for $d\mu$ which appears in the system.}
\begin{equation}
   p_2 \sim -p\ , \quad q_2 \sim \bar{\rho} \ ,\quad
   e^2_- \sim -z\ , \quad k_2 \sim - e_-\ , \quad \bar{w}_2 \sim - w\ , \quad w_2 \sim -\bar{w}\ .
\end{equation}
Intrinsic torsions will be fully reconstructed if equations (\ref{eq:de1e2basis}), (\ref{eq:de1e2cbasis}) and (\ref{eq:nablazbasis}) determines $p$, $q$ and $\rho$.

Equation (\ref{eq:nablazbasis}) is the only in which appears $\mathrm{Re}\,p$  ($\mathrm{Im}\,p$ appears in (\ref{eq:de1e2basis}) and in (\ref{eq:de1e2cbasis}) $p$ does not appear at all). So we conclude that $\mathrm{Re}\,p$ has to be determined by it. But we see, for example, that the component $w \cdot \mathrm{Re}\,p$ appears together with $z \cdot q$. Therefore this system can be sufficient only if we are able to obtain $z \cdot q$ in other places. But $z\cdot q$ is not determined in other places: indeed $z \cdot q$ is equivalent to $e_2^- \cdot q_1$, which is one of the components that are determined by the pairing equations. Therefore the system (\ref{eq:de1e2basis}), (\ref{eq:de1e2cbasis}) and (\ref{eq:nablazbasis}) is not sufficient to reconstruct the intrinsic torsions. To convince ourselves of this, we can compute the number of independent components which are present in (\ref{eq:de1e2basis}), (\ref{eq:de1e2cbasis}) and (\ref{eq:nablazbasis}). Starting with (\ref{eq:de1e2cbasis}) this is roughly given by the monomials
\begin{align}
  & \left[e_- \cdot q \right] + \left[ \bar{w} \cdot q \right] + \left[ w \cdot \bar{\rho} \right] + \left[ z \cdot \bar{\rho} \right] + \nonumber \\
  & + \left[ z \cdot \rho \right] + \left[ \bar{w} \cdot q \right] + \left[ w \cdot \bar{\rho} \right] + \left[ e_- \cdot q \right] \ .
\end{align}
Therefore we see that this expression determines $4$ complex components ($8$ real components). Moving to (\ref{eq:de1e2basis}), it takes the form
\begin{align}
  &  \left[ w \cdot \mathrm{Im}\,p +  e_- \cdot q \right] +\left[z \cdot \mathrm{Im}\,p + \bar{w} \cdot q \right] + \left[ w \cdot \rho \right] + \left[z \cdot \rho \right] + \nonumber \\
  & + \left[w \cdot \mathrm{Im}\,p + \rho \cdot z \right] + \left[w \cdot q \right] + \left[e_- \cdot \mathrm{Im}\,p + \bar{w} \cdot \rho \right] + \left[ e_- \cdot q \right]\ 
\end{align}
and in this manner we obtain $8$ real additional components. Finally, (\ref{eq:nablazbasis}) gives us the monomials
\begin{align}
  &  \left[e_- \cdot \mathrm{Re}\,p + z \cdot \mathrm{Re}\,p \right] + \left[z \cdot \mathrm{Re} p \right] + \left[\bar{w} \cdot \mathrm{Re}\,p + z \cdot \bar{q} \right] + \left[w \cdot \mathrm{Re}\,p + z \cdot q \right] + \nonumber \\
  & + \left[e_- \cdot \mathrm{Re}\,p \right] + \left[ e_- \cdot \bar{q} \right] + \left[e_- \cdot q \right] + \nonumber \\
  & + \left[w \cdot \mathrm{Re}\,p + e_- \cdot \rho \right] + \left[z \cdot \rho \right] + \left[w \cdot \bar{q} + \bar{w} \cdot \rho \right] + \left[ w \cdot q + w \cdot \rho \right] \nonumber \\
  & + \left[\bar{w} \cdot \mathrm{Re}\,p + e_- \cdot \bar{\rho} \right] + \left[ z \cdot \bar{\rho} \right] + \left[ \bar{w} \cdot \bar{q} + \bar{w} \cdot \bar{\rho} \right] + \left[\bar{w} \cdot q + w \cdot \bar{\rho} \right]
\end{align}
and so we obtain $6$ additional real components. Summarizing we see that (\ref{eq:de1e2basis}), (\ref{eq:de1e2cbasis}) and (\ref{eq:nablazbasis}) allow us to determine $22$ real components. These are not sufficient to fully reconstruct $p$, $q$ and $\rho$ which have $24$ real components. In conclusion we see that, in the timelike case as well, equations (\ref{eq:LK}) and (\ref{eq:psp10}) are not sufficient and at least one pairing equation is necessary.





\providecommand{\href}[2]{#2}


\end{document}